\documentclass[aps,
                pra,
                twocolumn,
                groupedaddress]{revtex4-2}
\usepackage{amsmath}
\usepackage{amssymb}
\usepackage{physics}
\usepackage{mathtools}
\usepackage{gensymb}
\usepackage{graphicx}
\usepackage{caption}
\usepackage{subcaption}
\usepackage{ragged2e} 
\usepackage{xcolor}
\usepackage[capitalise]{cleveref}
\usepackage{soul} 
\begin{document}


\title{Simulation of the Dissipative Dynamics of Strongly Interacting NV Centers with Tensor Networks}


\author{Jirawat Saiphet}
    \email[]{jirawat.saiphet@uni-tuebingen.de}
\author{Daniel Braun}
\affiliation{Institut f\"ur Theoretische Physik, Eberhard Karls Universit\"at T\"ubingen, 72076 T\"ubingen, Germany}


\date{\today}

\begin{abstract}
    NV centers in diamond are a promising platform for
    highly sensitive  quantum sensors for magnetic fields and other
    physical quantities.  The quest for high sensitivity
    combined with high spatial resolution leads naturally to
    dense ensembles
    of NV centers, and hence to strong, long-range interactions
    between them.  
    Hence, simulating strongly interacting NVs becomes
    essential. However, obtaining the exact dynamics for a many-spin
    system is a challenging task due to the exponential scaling of the
    Hilbert space dimension, a problem that is exacerbated when the
    system is modelled as an open quantum system.  
    In this work, we
    employ the Matrix Product Density Operator (MPDO) method to represent the
    many-body mixed state 
    and to simulate the
    dynamics of 
    an ensemble of NVs in the presence of strong
    long-range couplings due to dipole-dipole forces. We
    benchmark different time-evolution algorithms in terms of numerical
    accuracy and stability against time evolution based on exact numerical diagonalization. Subsequently, we simulate the dynamics in
    the strong interaction regime, and study the impact of decoherence on the accuracy of the MPDO method. 
    Lastly, we investigate the dynamics of quantum Fisher information and discuss under what circumstances a strong interaction can improve sensitivity for magnetic field sensing.
\end{abstract}


\maketitle

\section{Introduction}
Probing magnetic fields with high sensitivity and resolution is
important in frontier research applications. A single NV-center (NV
for short) in
diamond has been proposed as a nanoscale probe 
\cite{taylor_high-sensitivity_2008} and used for measuring a
magnetic field
\cite{welter_scanning_2022,dwyer_probing_2022}. Recently, controlled systems
of double and triple NV centers were successfully fabricated
\cite{haruyama_triple_2019}. Using NV ensembles with many spins has the potential to
increase the  sensitivity
by having more spins in a probe. 
\cite{balasubramanian_dc_2019,zheng_zero-field_2019,wang_realization_2022}. 
Requesting at the same time high
spatial resolution leads to  ensembles with high density. The
resulting strong dipole-dipole interactions between the NVs lead,
however, to a rapid population of sub-spaces of Hilbert space with
reduced total spin. E.g., if we had two spins-1/2, not only the triplet states would get populated, but also the anti-symmetric singlet state, resulting in a reduced response to the applied magnetic field. 
This can be considered a form of intrinsic
decoherence \cite{zhou_quantum_2020} in addition to the remaining external decoherence
mechanisms, resulting in short coherence times
\cite{bauch_decoherence_2020} and hence less sensitivity.
Decoupling the interactions with specific control pulses, or
engineering the alignments of NVs to reduce interactions can increase
the coherence time and sensitivity
\cite{zhou_quantum_2020,farfurnik_identifying_2018,osterkamp_engineering_2019}.
However, ideally one would
like to profit from
the interactions for generating entangled states that could highly
enhance the sensitivity of the probe. 

To model and optimize entanglement generation in a dissipative system with strong
and long-range interactions, simulations of its dynamics in the
presence of microwave control pulses is necessary. However, given the
many-body nature of an ensemble, simulating its exact dynamics is
intractable. Exact simulations of closed spin-1/2 systems with
long-range interaction have been implemented up to 32 spins
\cite{sandvik_ground_2010,schiffer_many-body_2019,zhou_quantum_2020,cheng_many-body_2023},
and up to 12 spins with dissipation \cite{essink_boundary-driven_2018,kucsko_critical_2018}. 
In order to address this challenge, we use a tensor network approach
to capture the dynamics of the ensemble. Matrix Product States (MPS)
\cite{vidal_efficient_2003,perez-garcia_matrix_2007,schollwock_density-matrix_2011,orus_practical_2014}
were proposed for the efficient simulation of quantum metrology
\cite{jarzyna_matrix_2013} and open quantum systems
\cite{finsterholzl_using_2020}. Simulation of non-Markovian systems has
been performed using Matrix Product Operators (MPO) with a
nearest-neighbors model
\cite{schlimgen_quantum_2022,fux_tensor_2023}. 

In this work, we consider an NV ensemble that consists of spin-1
particles. All NVs interact with each other by long-range
dipole-dipole interaction. We simulate the dissipative dynamics by using
the Matrix Product Density Operator (MPDO) method. We investigate the efficiency of
using MPDO in simulating dynamics in the strong interaction limit and
under dissipation. Operator entanglement entropy (opEE) is computed
and used to demonstrate the interplay between strong interaction and
dissipation for the capability of the MPDO to approximate the exact
states. We then address and quantify 
possible sensitivity improvements from NV-NV interactions during the time
evolution using quantum Fisher information. 

\section{Theory}
\subsection{Strongly interacting NVs}
An NV-center  is a spin-1 system that forms in a carbon lattice of
diamonds if two adjacent carbons are replaced by a nitrogen atom and a vacancy. Without an external field applied, the diamond structure creates
a Zero Field Splitting (ZFS). Energy levels corresponding to states $|0\rangle$ and $|\pm1\rangle$, with $m_{s}=0, \pm 1$, respectively, are separated by $D=2\pi\times 2870$ MHz. 

Diamond's crystalographic axes provide 4 possible orientations of the principal axes of the NV, see \cref{fig:4_axes}, separating them into different groups. E.g.\, the NV in [111] group has $\hat{z}$ parallel with a unit vector $\frac{1}{\sqrt{3}}(1,1,1)$. Different groups have different couplings to a magnetic field, which leads to different energy levels that can be distinguished by optically detected magnetic resonance (ODMR).

In an external magnetic field, the field component $B_{z}$ along the NV's principal axis creates an additional energy splitting between $|-1\rangle$ and $|+1\rangle$ proportional to $B_{z}$. This allows selective transitions from the ground state $|0\rangle$ to one of the two exited states by applying a microwave field at resonance frequency. Here we restrict ourselves to a transition to $|-1\rangle$.
\iftrue
\begin{figure}
    \centering
    \begin{minipage}{.5\columnwidth}
        \vspace*{\fill}
        \centering
        \includegraphics[width=0.9\linewidth]{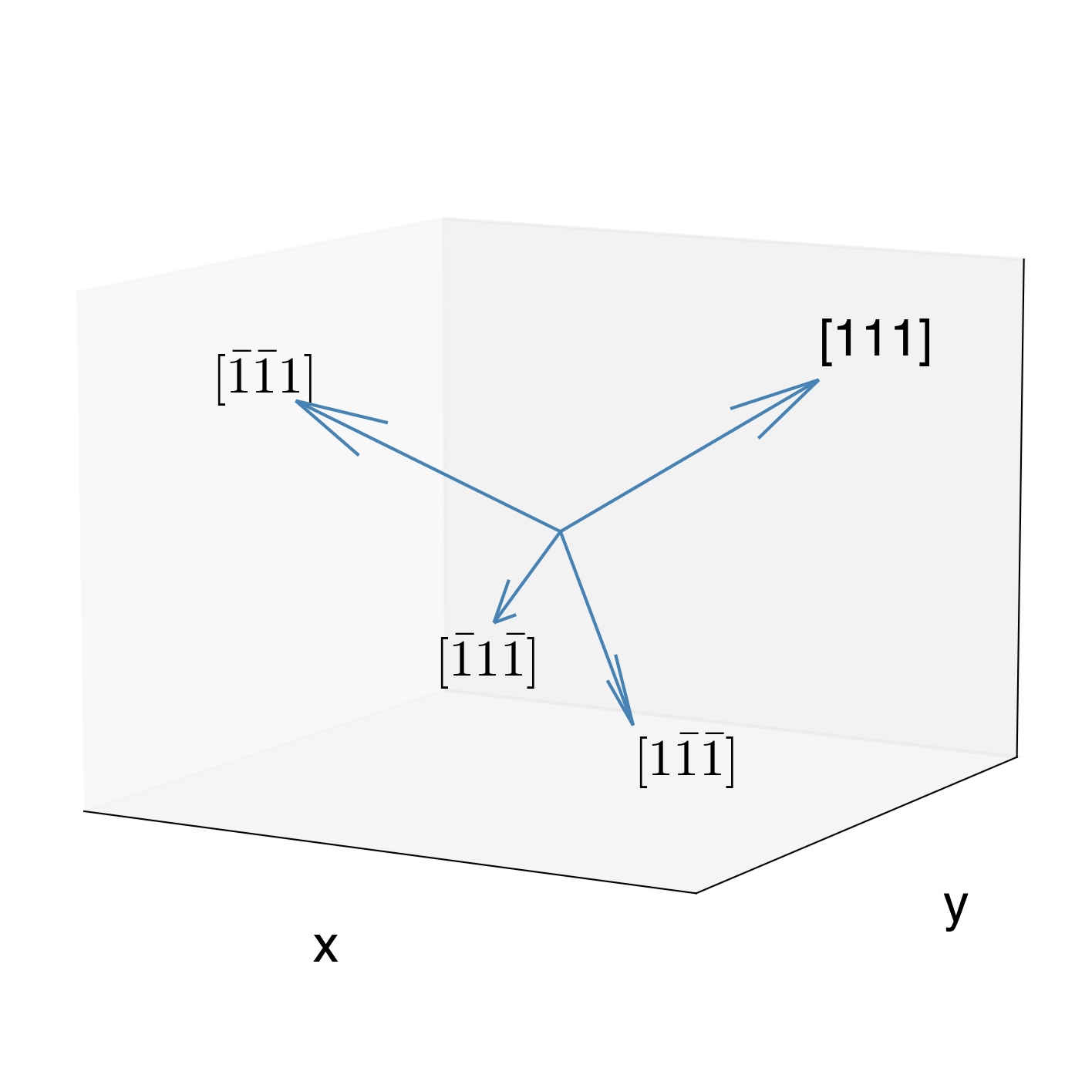}
        \subcaption{}
        \label{fig:4_axes}\par\vfill
        \includegraphics[width=0.9\linewidth]{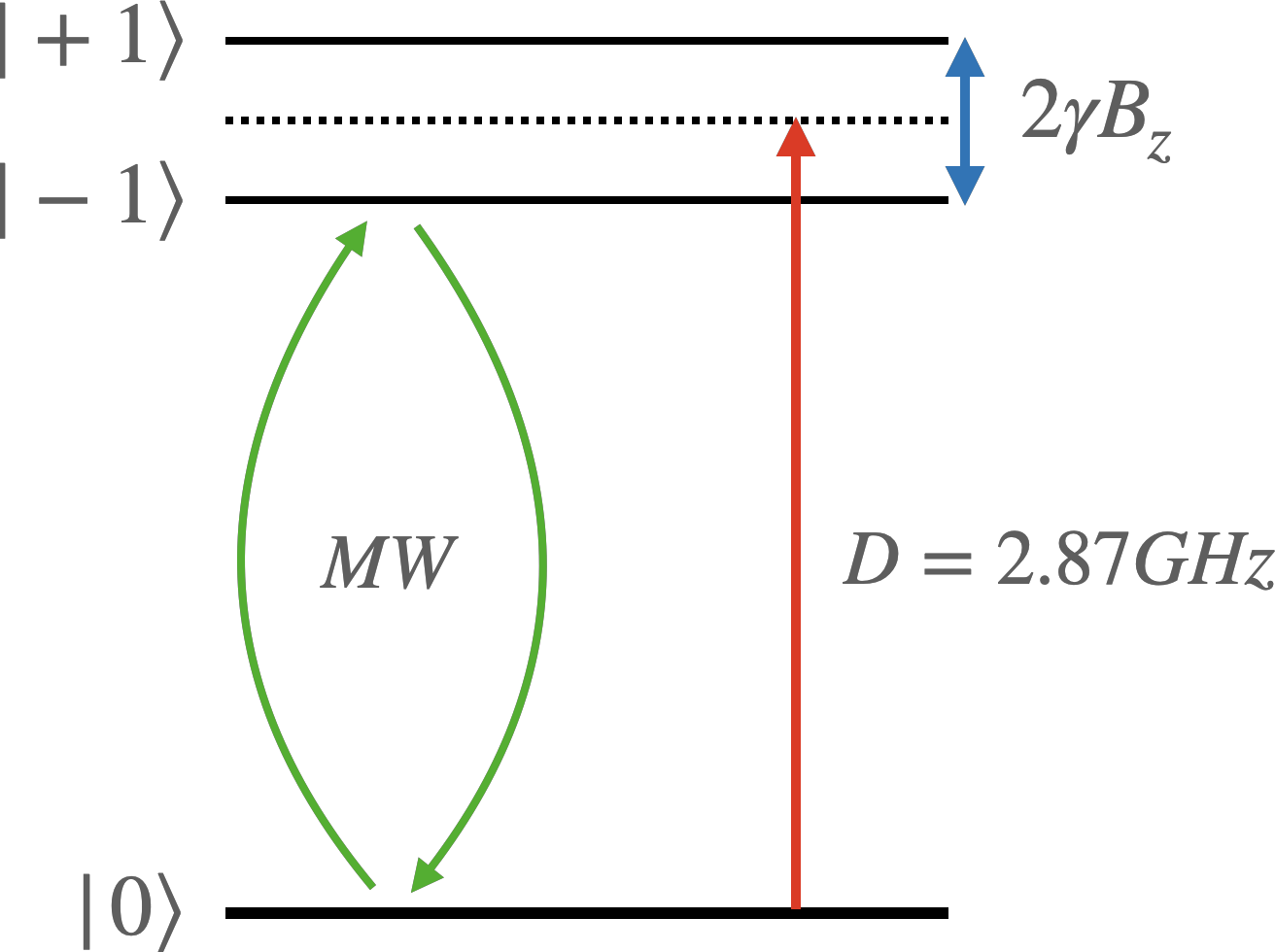}
        \subcaption{}
        \label{fig:spin-1}
    \end{minipage}%
    \begin{minipage}{.5\columnwidth}
        \includegraphics[width=\linewidth]{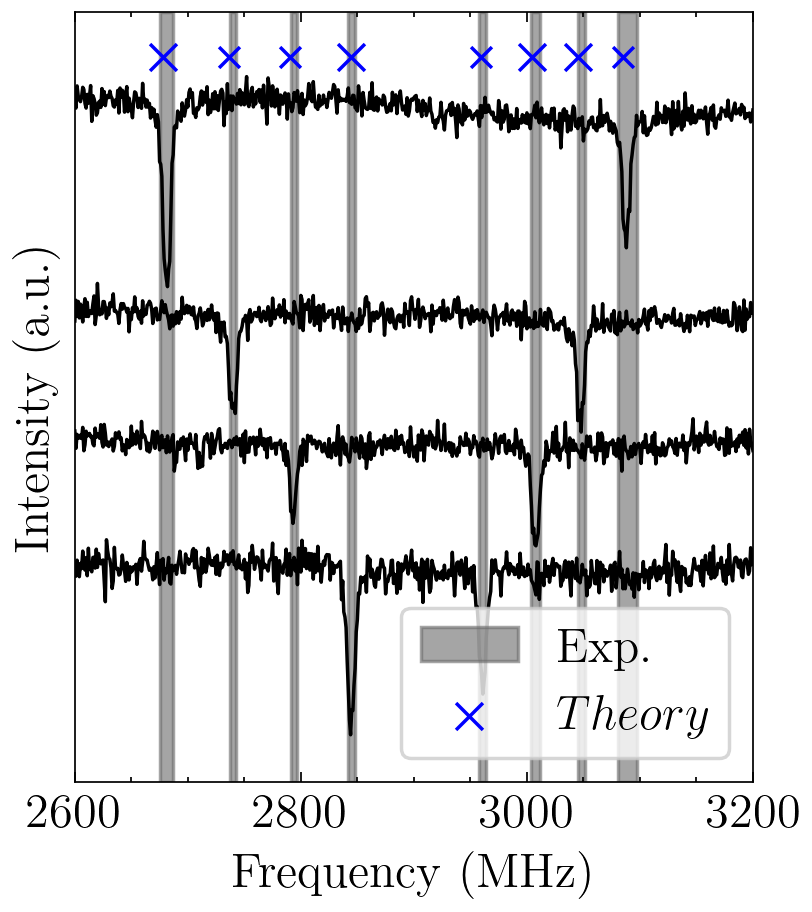}
        \subcaption{}
        \label{fig:spectrum_123}   
    \end{minipage}
    \caption{(\subref{fig:4_axes}) 4 possible orientations given by
      diamond's crystalographic axes. (\subref{fig:spin-1}) Energy levels of an NV
      center. (\subref{fig:spectrum_123}) ODMR
      spectra
      of 4 different axes from measurements
      and analytic calculations.}
\end{figure}
\fi

The Hamiltonian for an individual NV with microwave drive with Rabi frequency $\Omega(t)$ is given by
\begin{equation}
    \hat{H}_{NV,i} = \hbar\left( D\hat{S}_{z}^{2} + g_{s}\mu_{\text{B}} B^{(i)}_{z}\hat{S}_{z} + \Omega(t) \cos(\omega t)\hat{S}_{x}^{(i)} \right), \label{eqn:NVi}
\end{equation}
where $g_{s} \simeq 2$ and $\mu_\text{B}$ the Bohr magneton.
The spin operators for $S=1$ are 
\begin{align}
    \hat{S}_{x} = \frac{1}{\sqrt{2}}\begin{pmatrix}
                                        0& 1& 0\\
                                        1& 0& 1\\
                                        0& 1& 0
                                    \end{pmatrix},
    \hat{S}_{y} = \frac{i}{\sqrt{2}}\begin{pmatrix}
                                            0& -1& 0\\
                                            1& 0& -1\\
                                            0& 1& 0
                                        \end{pmatrix},\nonumber\\
    \hat{S}_{z} = \begin{pmatrix}
                    1& 0& 0\\
                    0& 0& 0\\
                    0& 0& -1
                \end{pmatrix}.
\end{align}
Applying the unitary $\hat{U}=e^{i\omega t\hat{S}_{z}^{2}}$ to
\cref{eqn:NVi} to transform to a frame co-rotating with the microwave 
and using rotating-wave approximation (RWA) yields
\begin{equation}
    \hat{H}_{\text{NV},i} = \hbar\left((D-\omega)\hat{S}_{z}^{2} + g_{s}\mu_{\text{B}} B^{(i)}_{z}\hat{S}_{z} + \frac{\Omega(t)}{2}\hat{S}_{x}^{(i)}\right) \label{eq:NVi_rwa}.
\end{equation} 
At resonance frequency, $\omega = D\pm g_{s}\mu_{\text{B}} B^{(i)}_{z}$, the microwave drive makes a transition between $|0\rangle \leftrightarrow |\pm 1\rangle$ for the $i$-th NV.
The energy levels of $\hat{H}_{\text{NV},i}$ are depicted in \cref{fig:spin-1}. They agree with the experimental observation \cite{haruyama_triple_2019}.

NVs interact with each other via dipole-dipole interaction. We consider a
case of strong interaction between NVs and hence ignore nuclear
spin. The general definition of dipole-dipole interaction between NV-$i$ and NV-$j$ is 
\begin{align}
    H_{\text{dip},ij} = -\frac{\mu_{0}(g_{s}\mu_{\text{B}})^{2}\hbar^{2}}{4\pi r_{ij}^{3}} (3(\vec{S}^{(i)}\dotproduct\hat{r})(\vec{S}^{(j)}\dotproduct\hat{r}) - \vec{S}^{(i)}\vec{S}^{(j)}), \label{Hdip}
\end{align} 
where $\vec{S}^{(i)} = (\hat{S}_{x}, \hat{S}_{y}, \hat{S}_{z})^{(i)}$ are spin operators of NV-$i
$, and $\hat{r}_{ij}=(r_{x},r_{y},r_{z})_{ij}$ are the unit vectors connecting the two NVs. We restrict ourselves to the case where all NV centers have the same orientation. The dipole-dipole interaction in the rotating frame can then be transformed into an effective Hamiltonian \cite{kucsko_critical_2018},
\begin{equation}
    \resizebox{0.48\textwidth}{!}{$
    \hat{H}_{\text{eff},\{ij\}} = 
        C_\text{dip}(\frac{1}{2}(\hat{S}_{x}^{(i)}\hat{S}_{x}^{(j)} + \hat{S}_{y}^{(i)}\hat{S}_{y}^{(j)}) - \hat{S}_{z}^{(i)}\hat{S}_{z}^{(j)}) \label{eqn:dip}$}
\end{equation}
where $C_\text{dip}=\frac{J_{0}q_{ij}}{r_{ij}^{3}}$, $J_{0}=\frac{\mu_{0}h^{2}(g_{s}\mu_{\text{B}})^{2}}{4\pi}=2\pi\times 52$\,MHz$\cdot$\,nm$^{3}$,
$q_{ij}=3(\hat{r}_{ij}\cdot\hat{z}_{i})(\hat{r}_{ij}\cdot\hat{z}_{j})
- \hat{z}_{i}\cdot\hat{z}_{j}$, with $\hat{z}_{i}$ the unit vector pointing in direction of dipole $i$. It reduces to $q_{ij}=(3\cos^{2}\theta
-1)$ and $\cos \theta \equiv \hat{z}_{i}\cdot \hat{r}_{ij}$ in the
case of the same group ($\hat{z}_{i}=\hat{z}_{j}$).

\subsection{Tensor network state}
Due to the
$r^{-3}$
scaling, the interactions within an ensemble of NVs extends beyond the nearest-neighbors and becomes long-range. These all-to-all interactions increase complexity and limit our capability to exactly simulate the system to only a few spins.  
\begin{figure}
    \centering
    \begin{subfigure}{0.23\textwidth}
        \centering
        \includegraphics[width=0.9\linewidth]{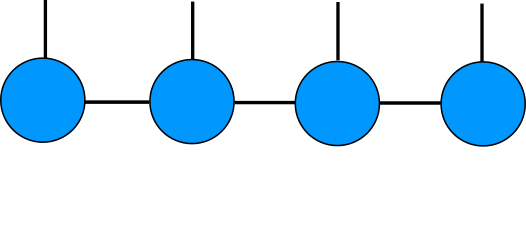}
        \caption[MPS]{} 
        \label{subfig:MPS}
    \end{subfigure}
    \begin{subfigure}{0.23\textwidth}
        \includegraphics[width=0.9\linewidth]{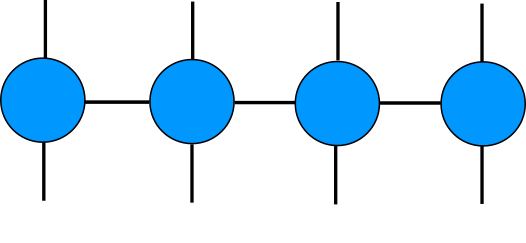}
        \caption[MPO]{}
        \label{subfig:MPO}
    \end{subfigure} \\
    \vspace{3ex}
    \begin{subfigure}{0.49\textwidth}
        \centering
        \includegraphics[width=\linewidth]{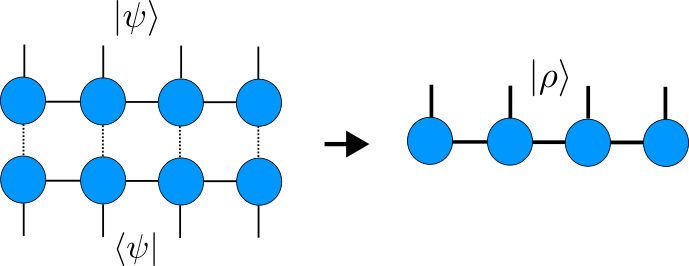}
        \caption[MPDO]{}
        \label{subfig:MPDO}
    \end{subfigure}
    \caption{Tensor diagrams representing (\subref{subfig:MPS}) MPS: a
      wavefunction for a pure state, (\subref{subfig:MPO}) MPO: an
      operator, and (\subref{subfig:MPDO}) MPDO: a vectorized density operator constructed from an MPS and its conjugate by contracting auxiliary indices, depicted as dash lines, and combining corresponding physical (bonds) indices. The thick line implies squared dimensions compared to the original index.}
    \label{fig:MPS/MPDO}
\end{figure}
To simulate the many-body dynamics for an ensemble of NVs, we
represent the quantum state as a tensor network state. The state of a closed system is decomposed into a
1-dimensional tensor network structure, called Matrix Product State
(MPS) \cite{vidal_efficient_2003,perez-garcia_matrix_2007,schollwock_density-matrix_2011,orus_practical_2014},

\begin{equation}
    |\Psi\rangle = \sum_{i_{1}...i_{N}=1}^{d} A^{i_{1}}A^{i_{2}}...A^{i_{N}}|i_{1}  \cdot\cdot\cdot i_{N}\rangle, \label{eq:mps}
\end{equation}
where a matrix $A^{i_{n}}$ has matrix elements $A^{i_{n}}_{\alpha_{(n-1)}\alpha_{n}}$. The MPS decomposition in \cref{eq:mps} can be visualized by a tensor diagram in \cref{subfig:MPS} where each sphere visualizes a tensor $A^{i_{n}}$ and lines are indices of the tensors. The connecting lines represent a contraction by summation over a shared index of two tensors. Each matrix elements $A^{i_{n}}_{\alpha_{(n-1)}\alpha_{n}}$ in the MPS
contains a physical index, $i_{n}$ representing the local Hilbert
space of the $n^{th}$ NV. The virtual index, or bond index $\alpha_{n}$
having bond dimensions $d(\alpha_{n})=\chi_{n}$ (except for $d(\alpha_{0}) = d(\alpha_{N}) = 1$) labels links between
two tensors. Those linked bonds will be contracted when we extract any observable from the MPS. For example, the contraction of a shared index between tensors $A^{i_{n-1}}A^{i_{n}} \equiv \sum_{\alpha_{(n-1)}=1}^{\chi_{(n-1)}} A^{i_{n-1}}_{\alpha_{(n-2)}\alpha_{(n-1)}}A^{i_{n}}_{\alpha_{(n-1)}\alpha_{n}}$ happens between the left bond index $\alpha_{n-1}$ of $A^{i_{n}}$ and the right bond index $\alpha_{n-1}$ of $A^{i_{n-1}}$. Note that in \cref{eq:mps} we use notation where the summation over the $\alpha_{n}$ are not written explicitly, and $n=1,\dotsc,N$. Physically, the bond dimension contains information about
the entanglement entropy between the two parts of the tensor network
that the bond connects. Using MPS allows us to compress the bond
dimensions and to efficiently represent the ground state in compact,
small Hilbert spaces.

A similar tensor network structure can be adapted to simulate an open system \cite{weimer_simulation_2019,jaschke_one-dimensional_2018}. To represent an operator we need a Matrix Product Operator (MPO) as given in \cref{subfig:MPO}. This MPO, in an orthogonal basis, represents a density operator of the system,
\begin{equation}
    \hat{\rho} = \sum_{i_{1}...i_{N}=1}^{d}\sum_{i'_{1}...i'_{N}=1}^{d} B^{i_{1}i'_{1}}B^{i_{2}i'_{2}}...B^{i_{N}i'_{N}}|i_{1}  \cdot\cdot\cdot i_{N}\rangle\langle i'_{1}  \cdot\cdot\cdot i'_{N}|.   \label{eq:mpdo1}
\end{equation}
For pure states, this MPO can be constructed by contracting auxiliary indices of two MPS and then combining corresponding physical (bond) indices. Furthermore, we combine the physical indices to create a Matrix Product Density Operator (MPDO) representing a vectorized density operator.
\begin{equation}
    |\rho\rangle\rangle = \sum_{j_{1}...j_{N}=1}^{d^{2}} B^{j_{1}}B^{j_{2}}...B^{j_{N}}|j_{1}  \cdot\cdot\cdot j_{N}\rangle\rangle.  \label{eq:mpdo2}
\end{equation}
Here, $j_{n}=(i_{n},i'_{n})$ and $|j_{n}\rangle\in\mathbb{C}^{d^{2}}$
is combined from two physical indices. The dimension of each resulting
index is doubled compared to the MPS. A graphical representation of this process is shown in \cref{subfig:MPDO}

\subsection{Simulation of dissipative dynamics}
We simulate $\rho$ directly, including dissipation, by solving the vectorized master equation
\begin{equation}
    \frac{\partial}{\partial t}|\rho\rangle\rangle =\mathcal{L}|\rho\rangle\rangle,
\end{equation} 
where $|\rho\rangle\rangle$ is the MPDO given by \cref{eq:mpdo2}, and $\mathcal{L}$ is a vectorized Lindblad operator defined as 
\begin{align}
    \mathcal{L}(t) = &-i
(\hat{H}(t)\otimes \mathbb{I} - \mathbb{I} \otimes \hat{H}^{T}(t)) \nonumber\\
&+ \sum_{i} \gamma_{i}[L_{i}\otimes (L^{\dagger}_{i})^{T} - \frac{1}{2}(L^{\dagger}_{i}L_{i}\otimes \mathbb{I} +  \mathbb{I}\otimes (L^{\dagger}_{i}L_{i})^{T})].
\end{align}
Note that when $\gamma_{i}=0$, the dynamics are unitary. In this case when MPS is sufficient to simulate the system, utilizing MPDO is unnecessary and computationally more expensive due to the squared memory. 

\section{Results}
We simulate a 1-dimensional spin-1 chain with
long-range interaction and dissipation using MPDO. We assume uniform separation for nearest neighbor spins,
$r_{i,i+1}=r$ for any $i=1,2,\dotsc,N$. All spins align on the $xy$-plane such that the position $\vec{r}^{(i)}=r(i,i,0)/\sqrt{2}$. For the very strong interaction regime, this separation is set
to be $r<2$\,nm while experiments with dense samples have reached $r \sim 5$\,nm
\cite{haruyama_triple_2019}. In our simulations we use the following
conditions: 
{\em i}) The amplitude of external magnetic field is chosen to match the level splitting read-off from the top line in Fig.~\ref{fig:spectrum_123}, $2g_s\mu_BB_z=2\pi\times 407$\,MHz. 
{\em ii}) Uniform Rabi frequency $\Omega(t)=\Omega=2\pi\times 2.00$ MHz  for driving the $|0\rangle \leftrightarrow |-1\rangle$ transition, $\omega = D - g_{s}\mu_{B} B_{z}$. 
{\em iii}) All NVs belong to the same orientation group parallel to [111]. 
({\em iv}) Time step $dt=1$ ns.
({\em v}) Initial state = $|0\rangle^{\otimes N}$.

\subsection{Simulation algorithms in strong interaction regime without dissipation}
\begin{figure}[h]
    \subfloat[]{%
    \includegraphics[width=.49\linewidth]{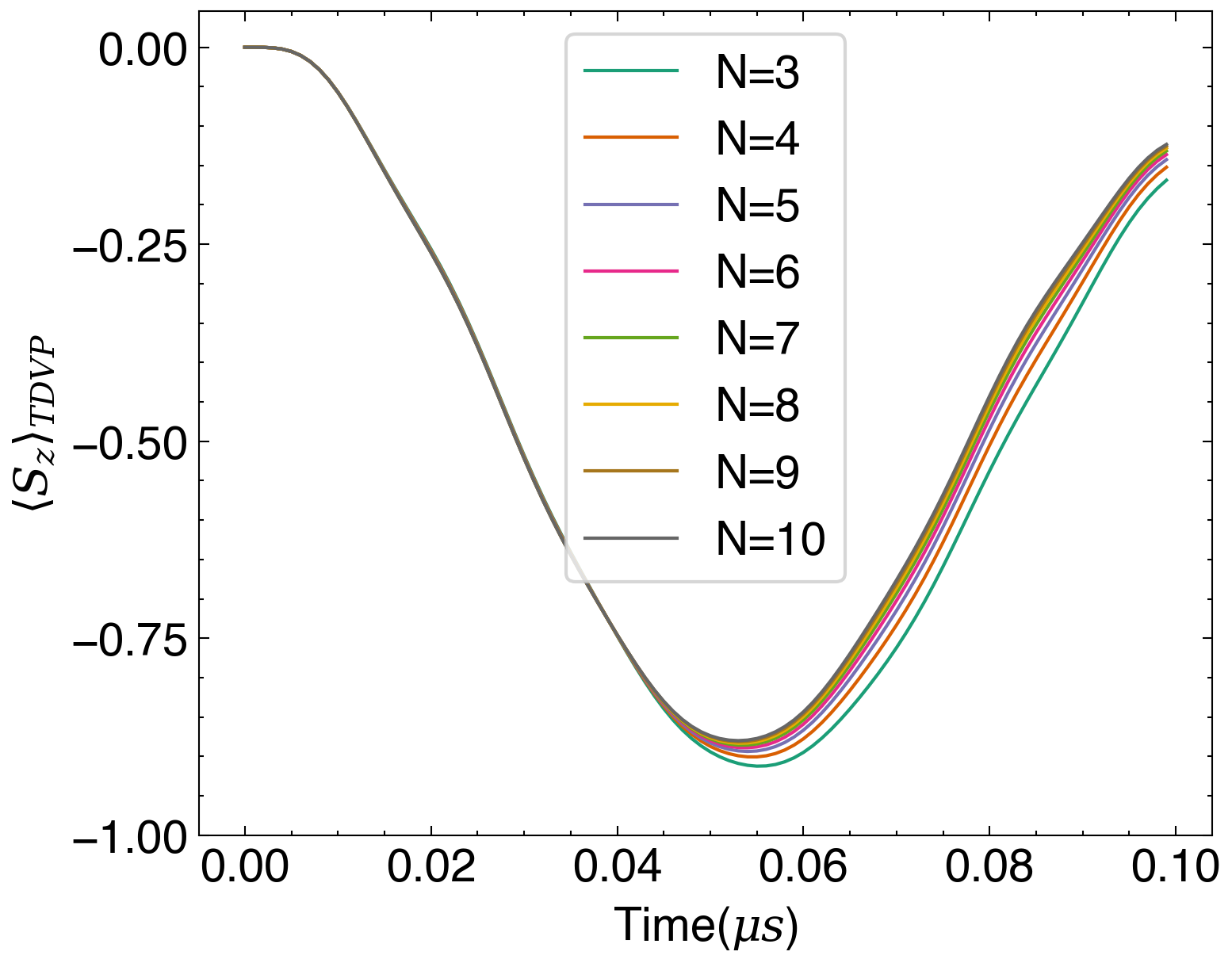}%
    \label{subfig:EngineComparison_0}%
    }\hfill
    \subfloat[]{%
        \includegraphics[width=.49\linewidth]{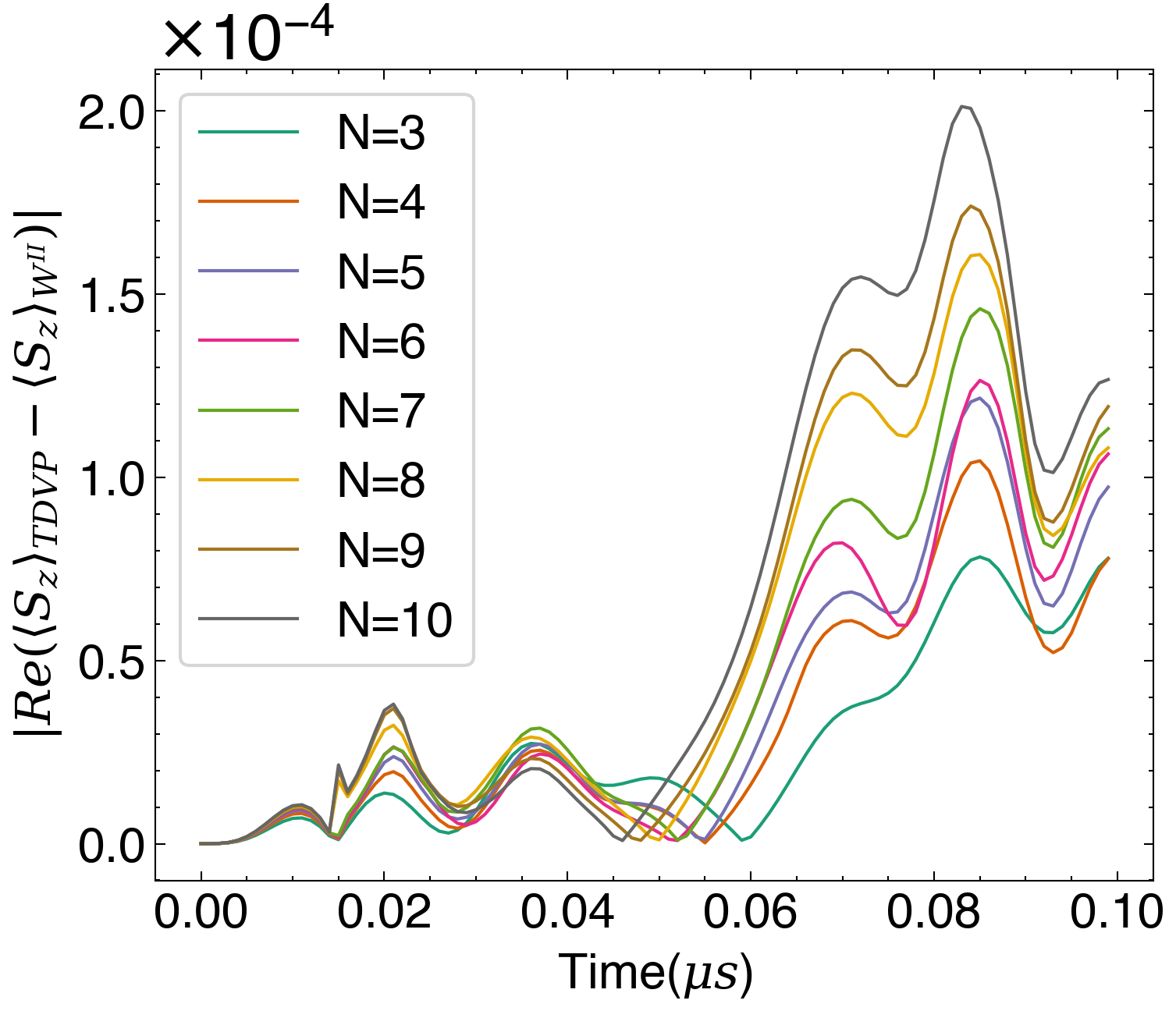}%
        \label{subfig:EngineComparison_a}%
    }\hfill
    \subfloat[]{%
        \includegraphics[width=.49\linewidth]{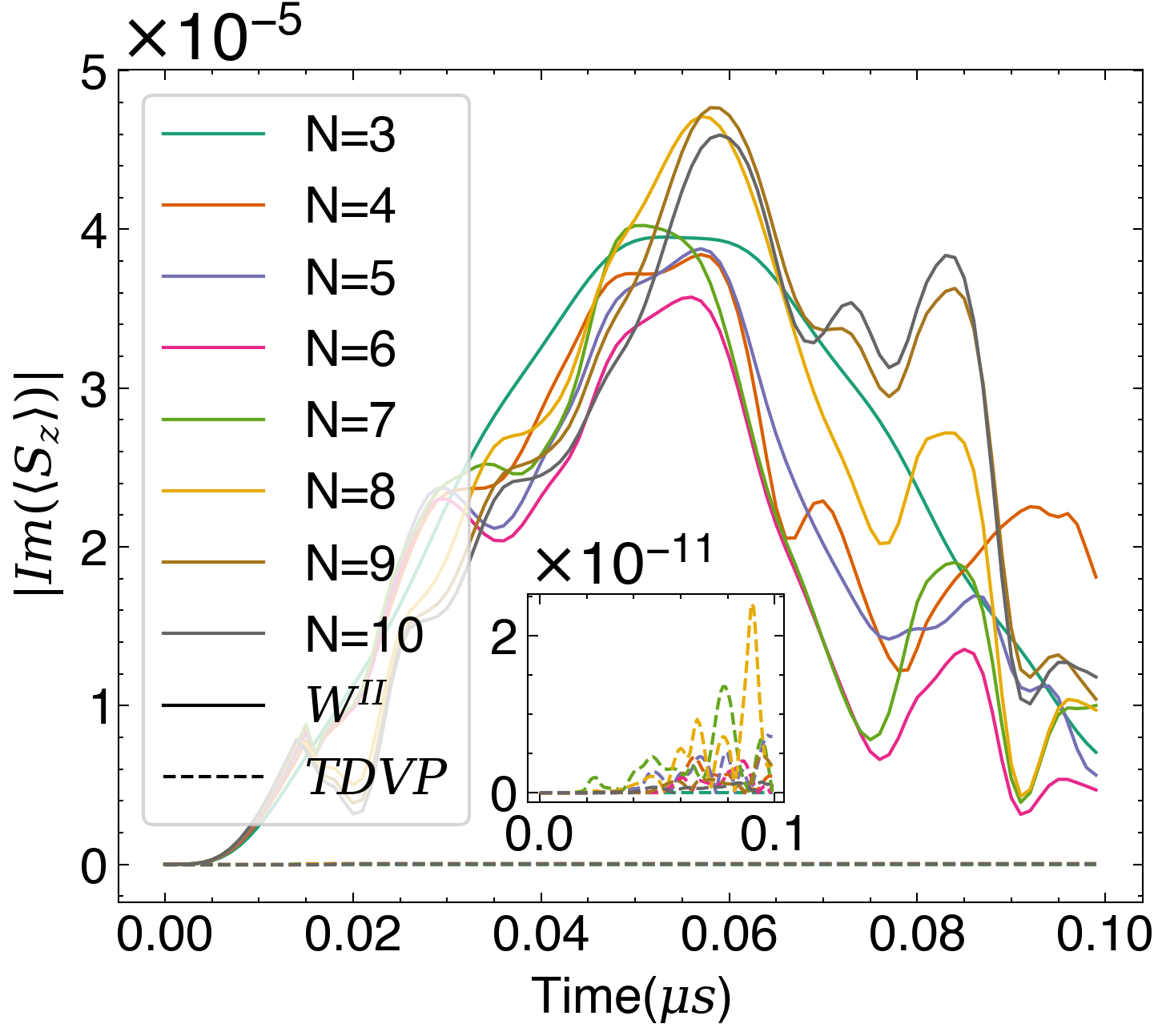}%
        \label{subfig:EngineComparison_b}%
    }
    \caption{Comparison of simulation results between $W^{II}$ and TDVP algorithms for different $N$.
    (\subref{subfig:EngineComparison_0}) Real-time evolution of $\langle S_{z} \rangle$ calculated from TDVP algorithm with MPDO.
    (\subref{subfig:EngineComparison_a}) difference in real part of $\langle S_{z}\rangle$ from TDVP and $W^{II}$ algorithms. 
    (\subref{subfig:EngineComparison_b}) Numerical errors in imaginary
    part of $W^{II}$ algorithm (main plot) are 5 orders of magnitude larger than in the TDVP algorithm (inset).}
    \label{fig:EngineComparison}
\end{figure}

\begin{figure}
    \centering
    \includegraphics[width=.98\linewidth]{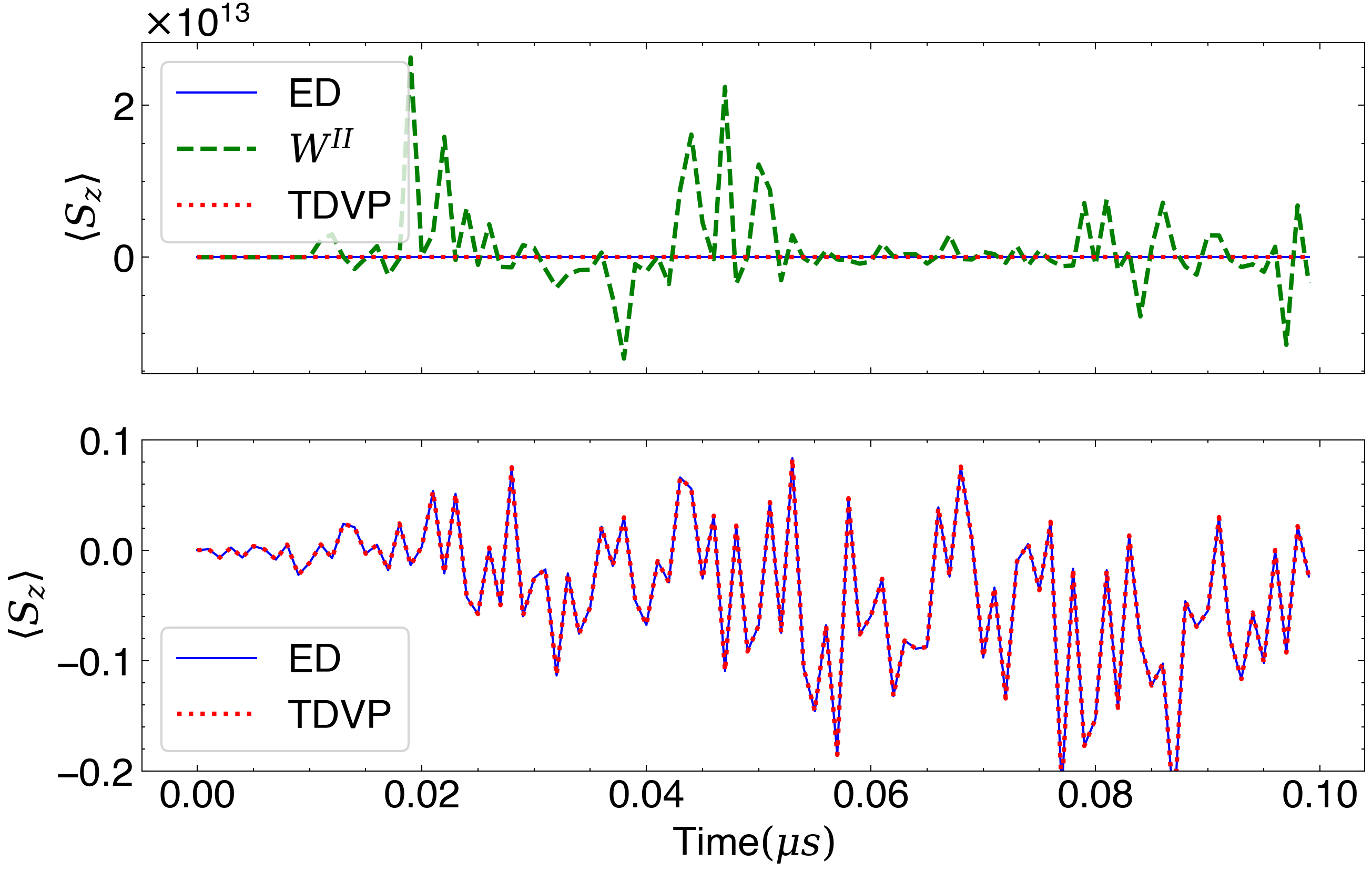}%
    \caption{ $W^{II}$ can fail when interaction becomes very strong (see top panel) while TDVP gives the same dynamics as exact diagonalization (bottom panel). Results from $N=3$ and $r=0.5$\,nm.}
    \label{subfig:EngineComparison_c}%
\end{figure}

Firstly, we investigate the precision and numerical stability of tensor network algorithms in the very strong interaction regime with $\gamma_{i}=0$, for two different algorithms that simulate time-evolution and support the long-range model \cite{paeckel_time-evolution_2019}: i) MPO $W^{II}$ \cite{zaletel_time-evolving_2015} and ii) time-dependent variational principle (TDVP) \cite{haegeman_unifying_2016,hubig_error_2018}. We simulate dynamics of ensembles up to 10 NVs and $\chi_{\text{max}}=16$. Here $\chi_\text{max}$ is a maximum dimension for all bonds. First of all, we investigate that the same dynamics are produced from both $W^{II}$ and TDVP. This is shown as small discrepancy in the real part of $\langle S_{z}\rangle$ of the whole system produced by the two algorithms in \cref{subfig:EngineComparison_a} when $r=2$\,nm. However, the TDVP has significantly smaller errors in the imaginary parts, as shown in \cref{subfig:EngineComparison_b}.
Furthermore, in \cref{subfig:EngineComparison_c} we observe that the $W^{II}$ can be numerically instable under certain condition i.e.\  when the interaction Hamiltonian is fully modeled by \cref{Hdip} and not being approximated by \cref{eqn:dip}. The algorithm fails to produce a normalized value for expected magnetization $\langle S_{z} \rangle$ when the interaction strength becomes stronger while the TDVP is still valid compared to exact diagonalization. This is shown in \cref{subfig:EngineComparison_c} for $N=3$ and $r=0.5$\,nm. 
We suspect that the observable divergence in $W^{II}$ is a result of exploiting the complex time steps technique mentioned in \cite{paeckel_time-evolution_2019} to improve the propagation error. As using the technique destroys unitarity of the propagator. However, without the technique, the numerical error of propagating the quantum state with $W^{II}$ method compared to the exact evolution dramatically degrades from $\mathcal{O}(dt^{2})$ to $\mathcal{O}(dt)$. In this case, the algorithm would not be able to capture the correct dynamics even in the weaker interaction regime. To avoid this problem, we keep the TDVP as the main algorithm for our simulations.

\subsection{Dissipative dynamics with finite bond dimension}
\begin{figure}[h]
    \subfloat[]{%
        \includegraphics[width=.48\linewidth]{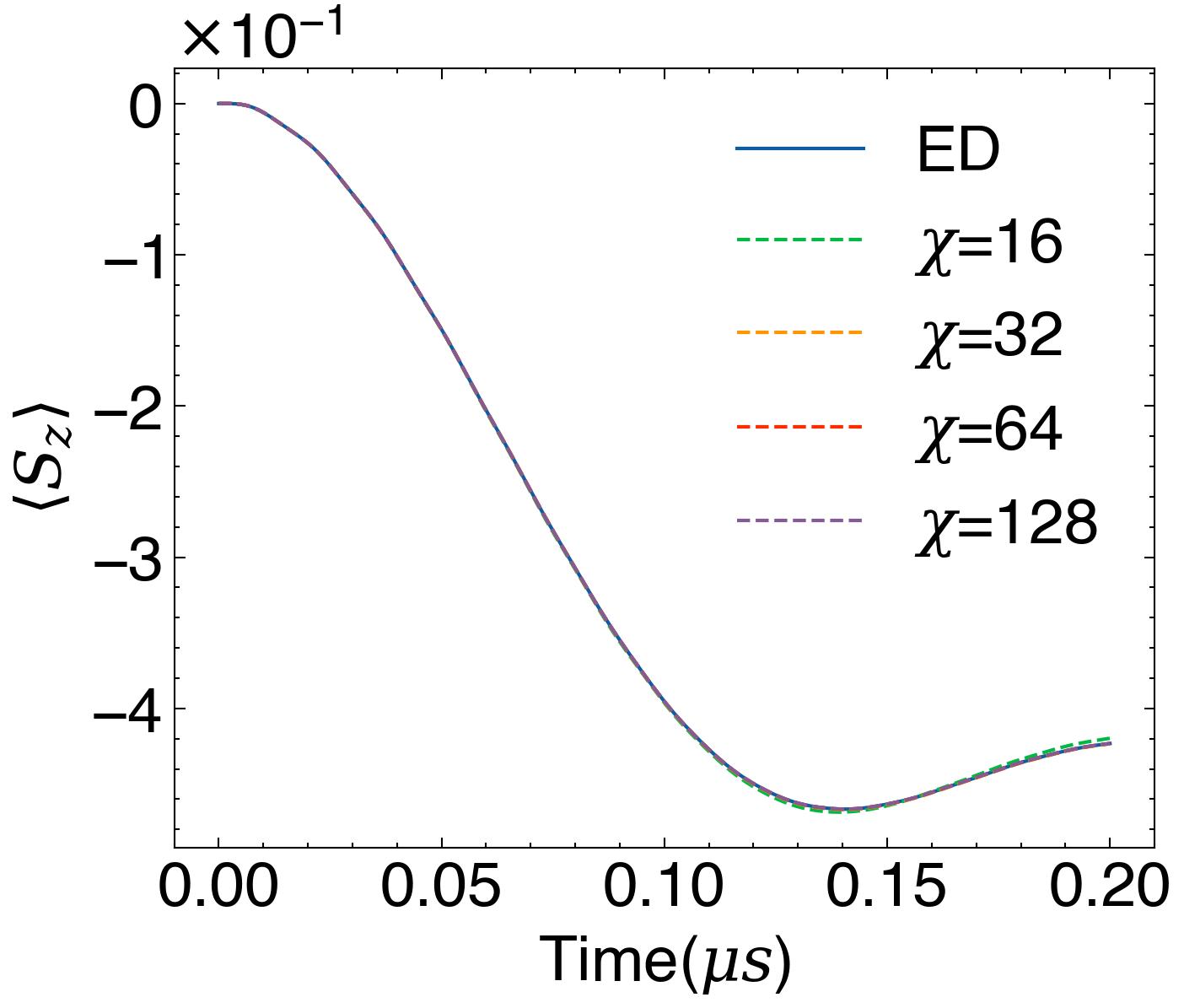}%
        \label{subfig:4NV_a}%
    }\hfill
    \subfloat[]{%
        \includegraphics[width=.48\linewidth]{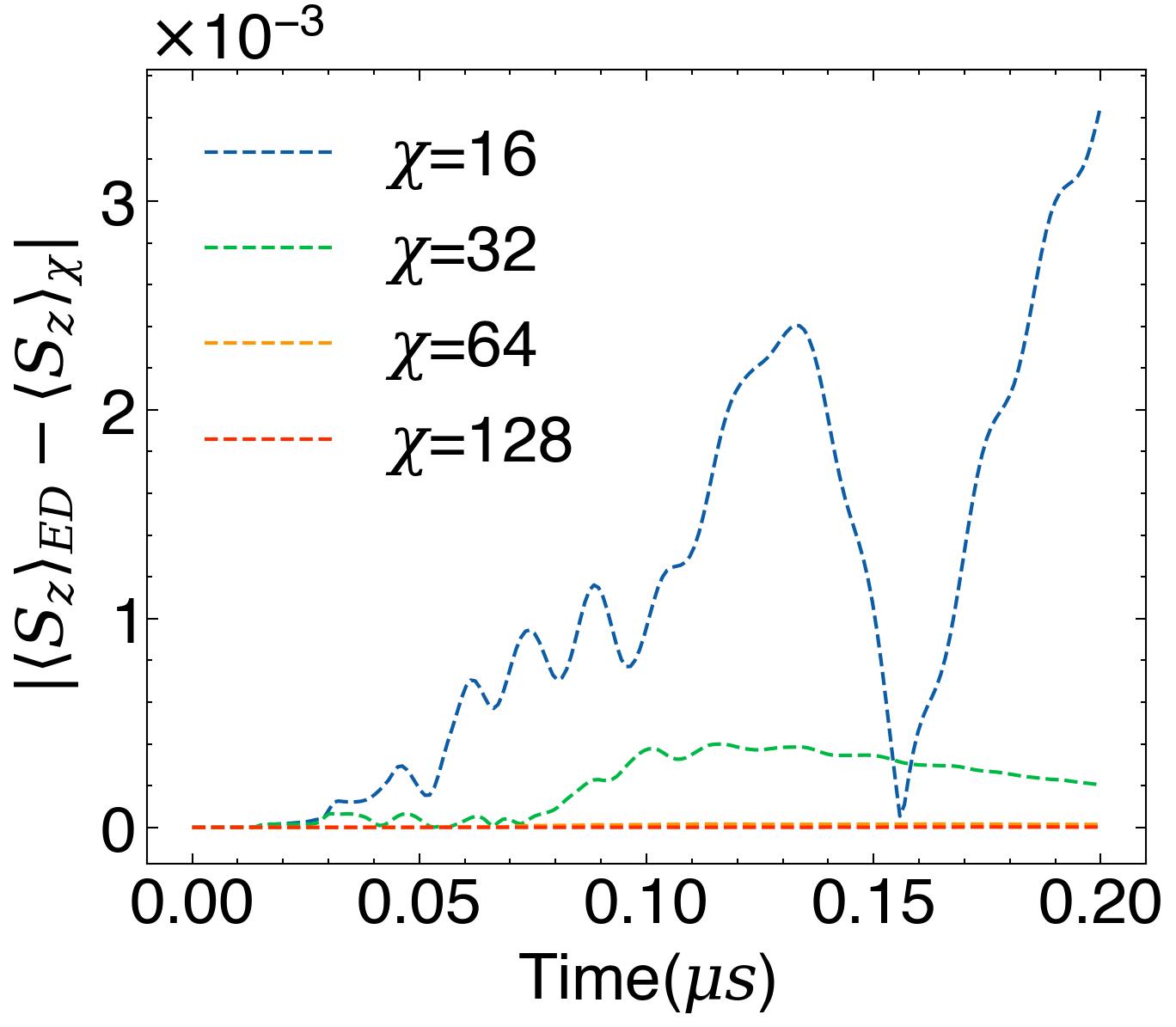}%
        \label{subfig:4NV_b}%
    }\\
    \subfloat[]{%
        \includegraphics[width=.48\linewidth]{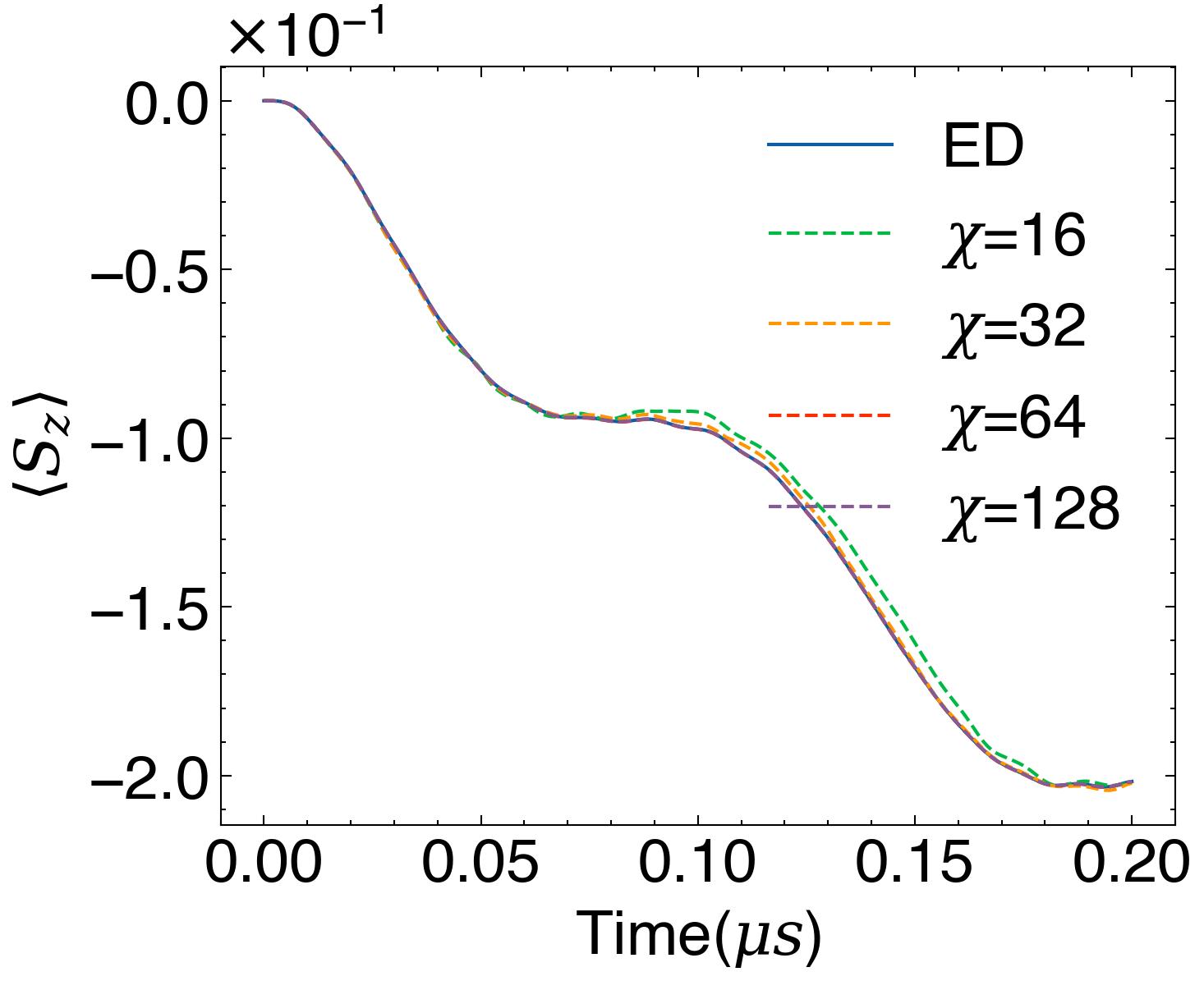}%
        \label{subfig:4NV_c}%
    }\hfill
    \subfloat[]{%
        \includegraphics[width=.48\linewidth]{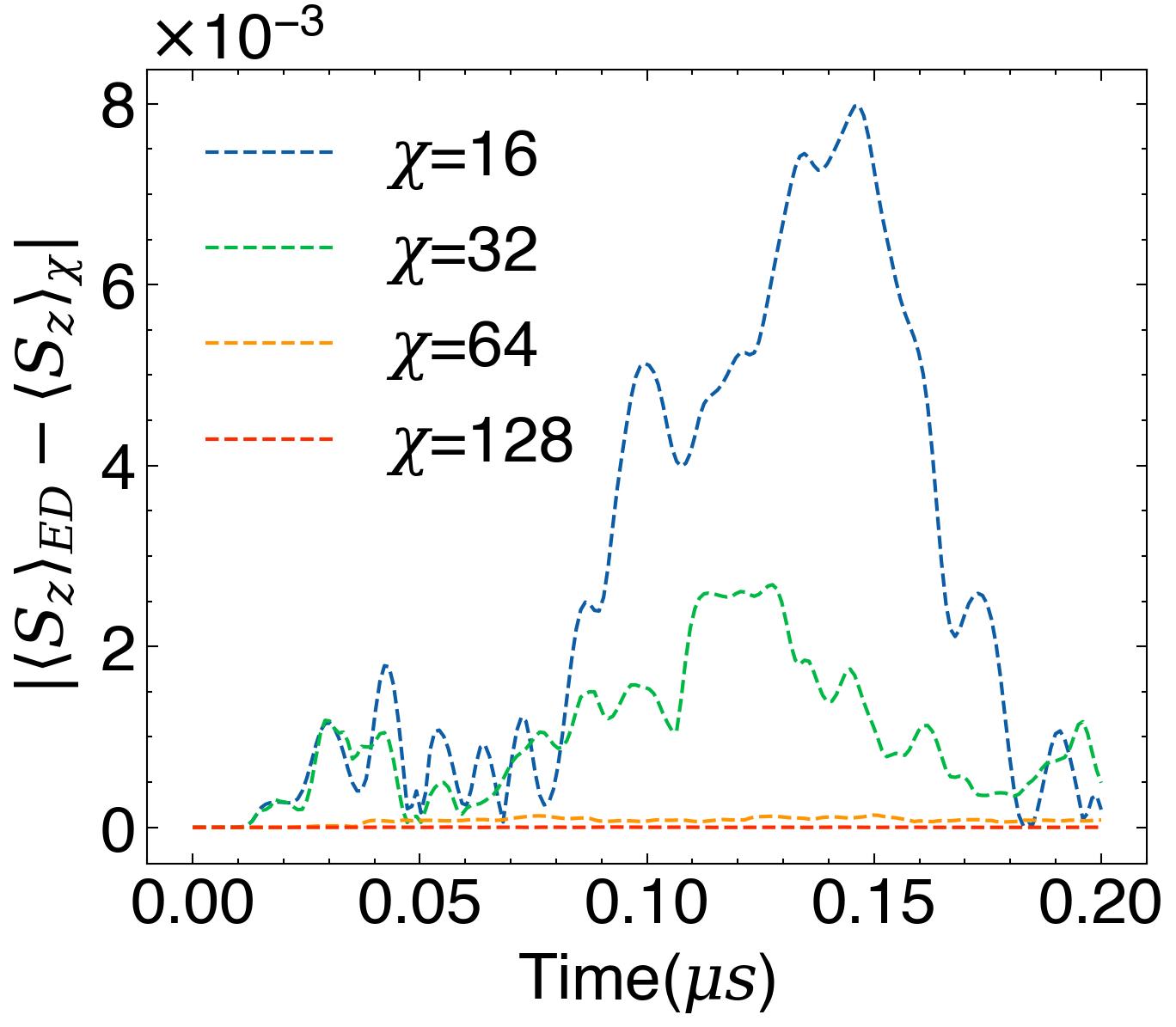}%
        \label{subfig:4NV_d}%
    }
    \caption{4NVs with $\gamma=0$. (\subref{subfig:4NV_a}) Average magnetization extracted from MPDO as a function of time and (\subref{subfig:4NV_b}) numerical errors compared to exact diagonalization as a function of maximum bond dimension for $r=2.0$\,nm. (\subref{subfig:4NV_c}), (\subref{subfig:4NV_d}) for $1.5$\,nm}
    \label{fig:4NV}
\end{figure}

\begin{figure}[h]
    \subfloat[]{%
        \includegraphics[width=.48\linewidth]{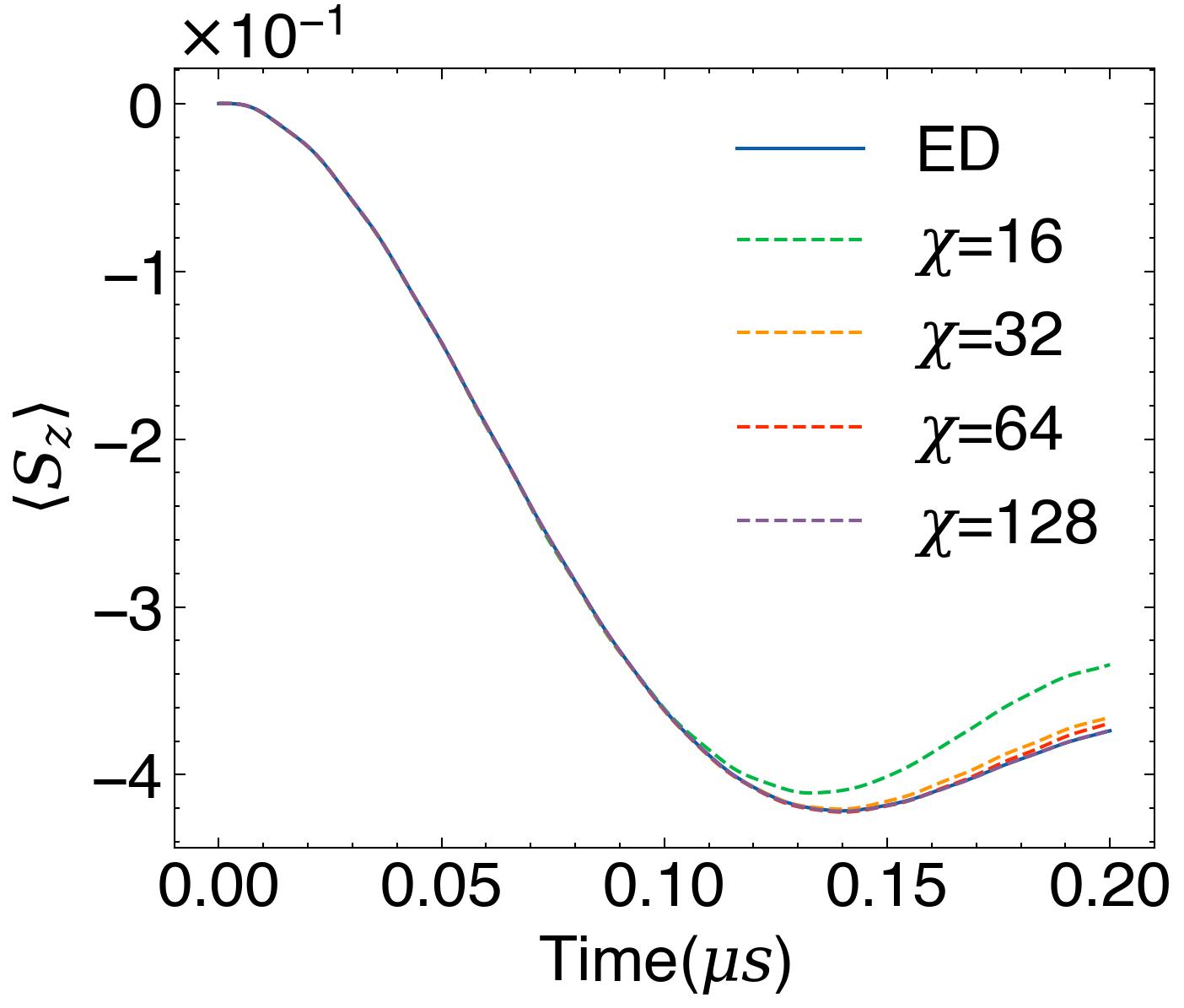}%
        \label{subfig:7NV_a}%
    }\hfill
    \subfloat[]{%
        \includegraphics[width=.48\linewidth]{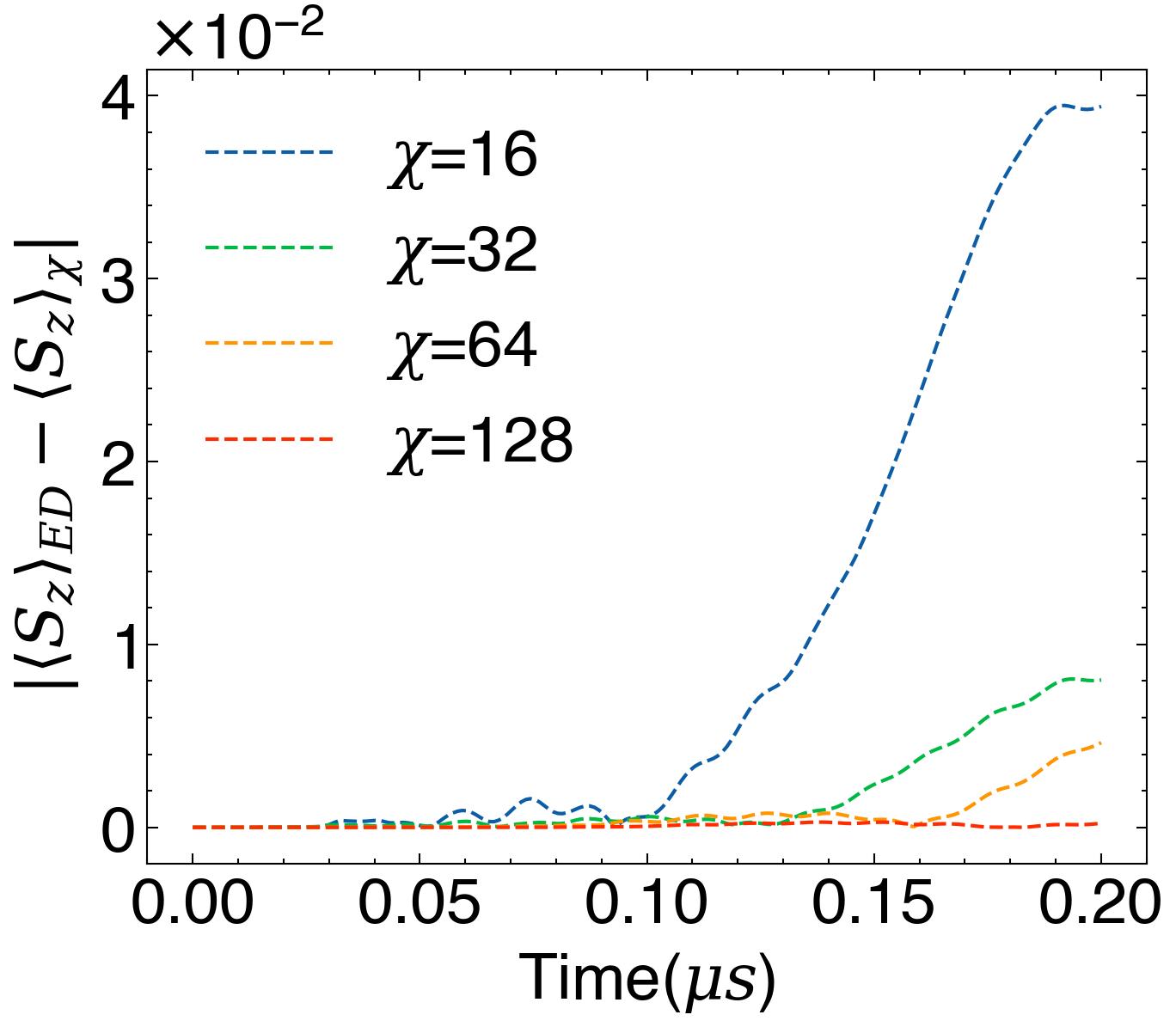}%
        \label{subfig:7NV_b}%
    }\\
    \subfloat[]{%
        \includegraphics[width=.48\linewidth]{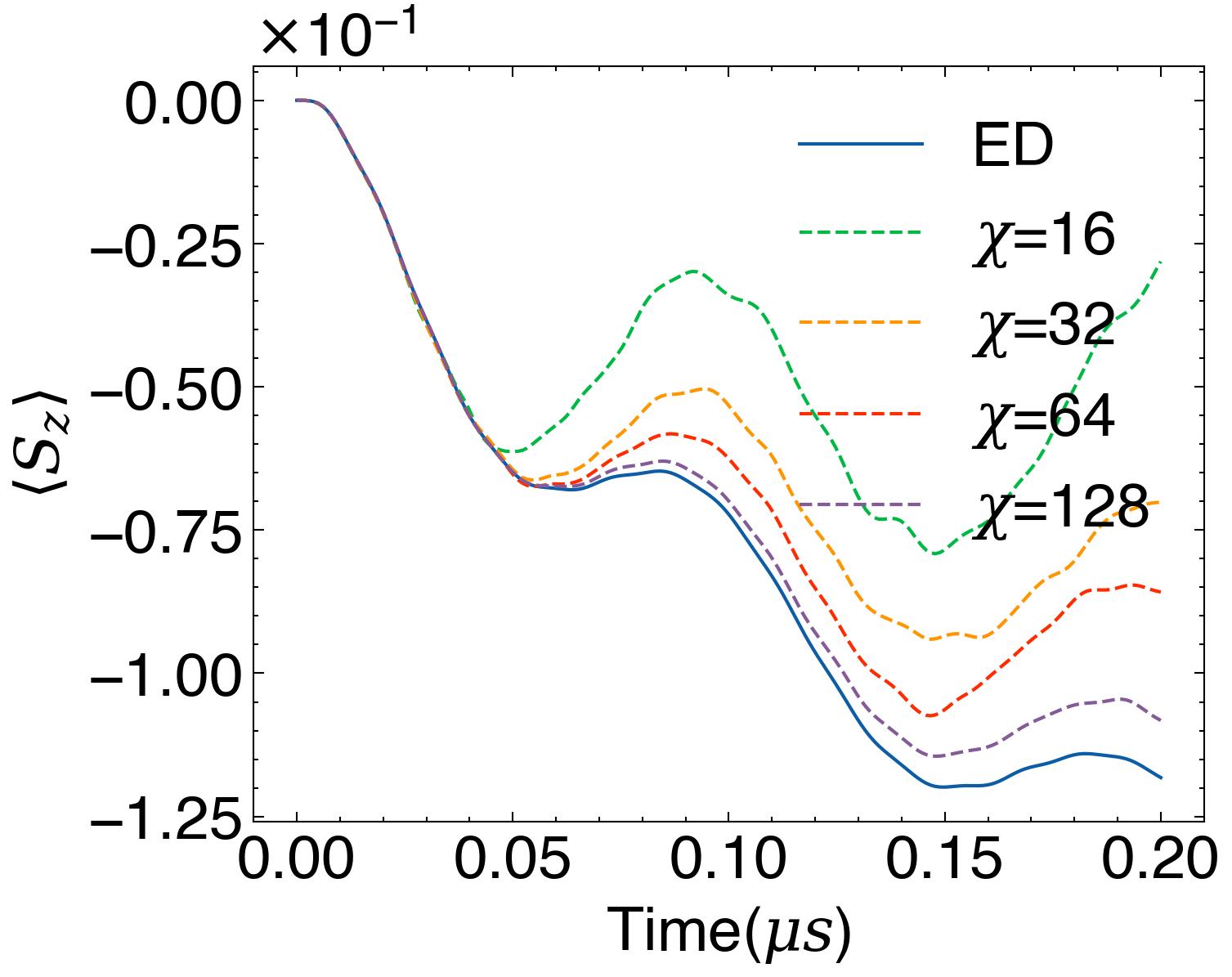}%
        \label{subfig:7NV_c}%
    }\hfill
    \subfloat[]{%
        \includegraphics[width=.48\linewidth]{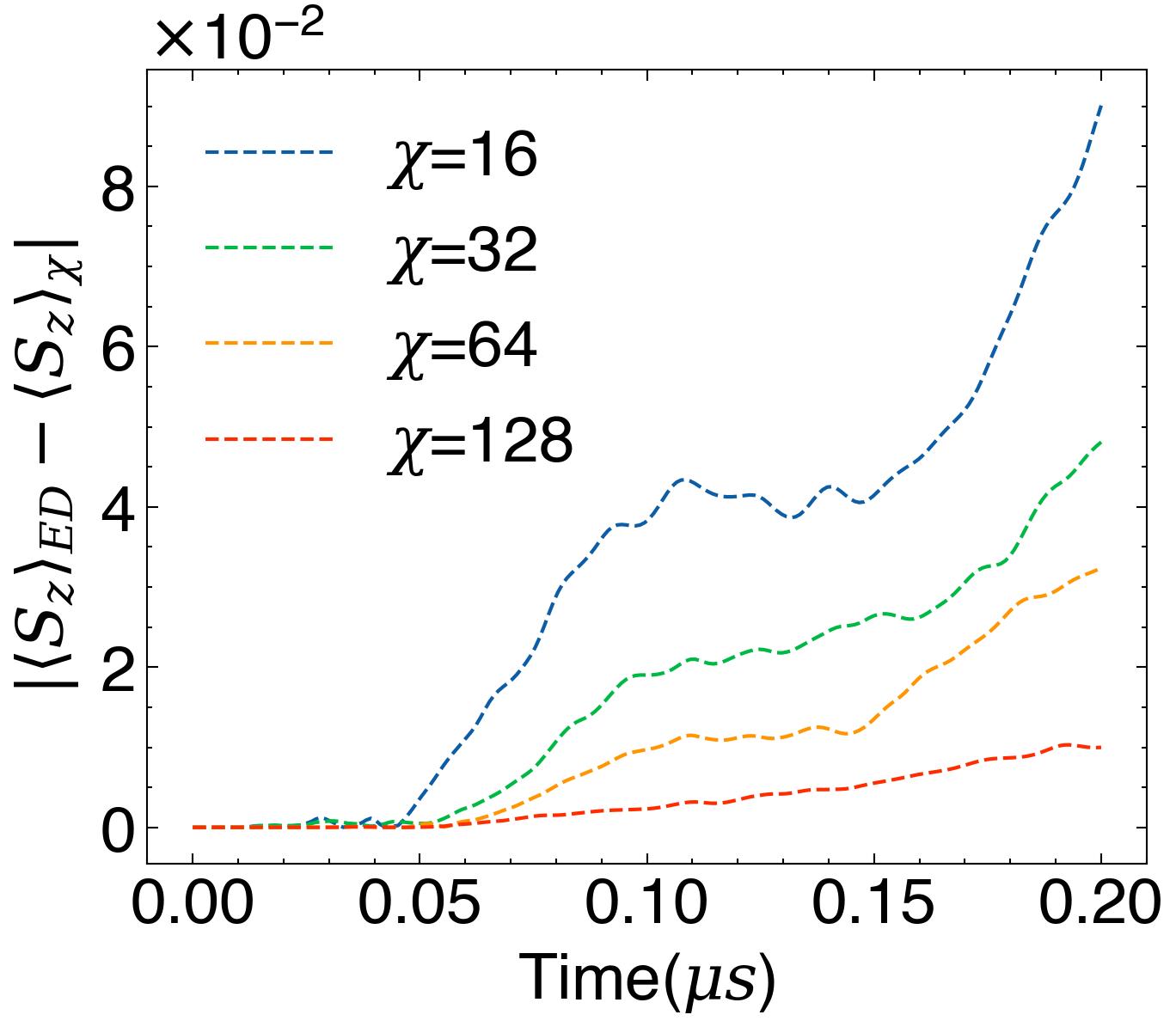}%
        \label{subfig:7NV_d}%
    }
    \caption{7NVs with $\gamma=0$. (\subref{subfig:7NV_a}) Average magnetization extracted from MPDO as a function of time and (\subref{subfig:7NV_b}) numerical errors compared to exact diagonalization as a function of maximum bond dimension for $r=2.0$\,nm. (\subref{subfig:7NV_c}), (\subref{subfig:7NV_d}) for $1.5$\,nm}
    \label{fig:7NV}
\end{figure}

A major source of error when using MPS/MPDO is truncation
error. Keeping finite bond dimensions and truncating when the bond
indices exceed $\chi_{\text{max}}$ introduces an error. This error is
given by the square root of the sum of squares of the truncated
singular values of all bonds,  
\begin{equation}
    \epsilon = \sqrt{\sum_{k>\chi_{\text{max}}} (s_{k})^{2}}\,.
\end{equation}
Typically, these errors
remain bounded
if the state has entanglement that follows an area-law, e.g.~when
the
system has only nearest-neighbor interactions. In such a case, the
bond
dimension
grows with small singular values during the time evolution, allowing
the simulation of large systems with small error. However, since we
are considering a case of long-range interaction, this argument should
not hold. In \cref{fig:4NV} and \cref{fig:7NV} we plot simulation
results for different $\chi_{\text{max}}$ with $r=2.0$\,nm and
$r=1.5$\,nm,
compared to a result from exact diagonalization for $N=4,7$. We
find that stronger interactions introduce bigger errors in bond
truncation. The results for smaller $\chi_{\text{max}}$ only follow
the exact calculation as long as entanglement entropy does still not
reach $\chi_{\text{max}}$;
i.e.~
up to a
certain number of
timesteps before diverging and becoming inaccurate due to truncation errors. 

These results imply that it requires bigger $\chi_{\text{max}}$ to
capture the dynamics in the presence of  long-range, strong
interactions, thus less efficiency of the MPS/MPDO method.

\begin{figure}[h]
    \subfloat[]{%
        \includegraphics[width=.48\linewidth]{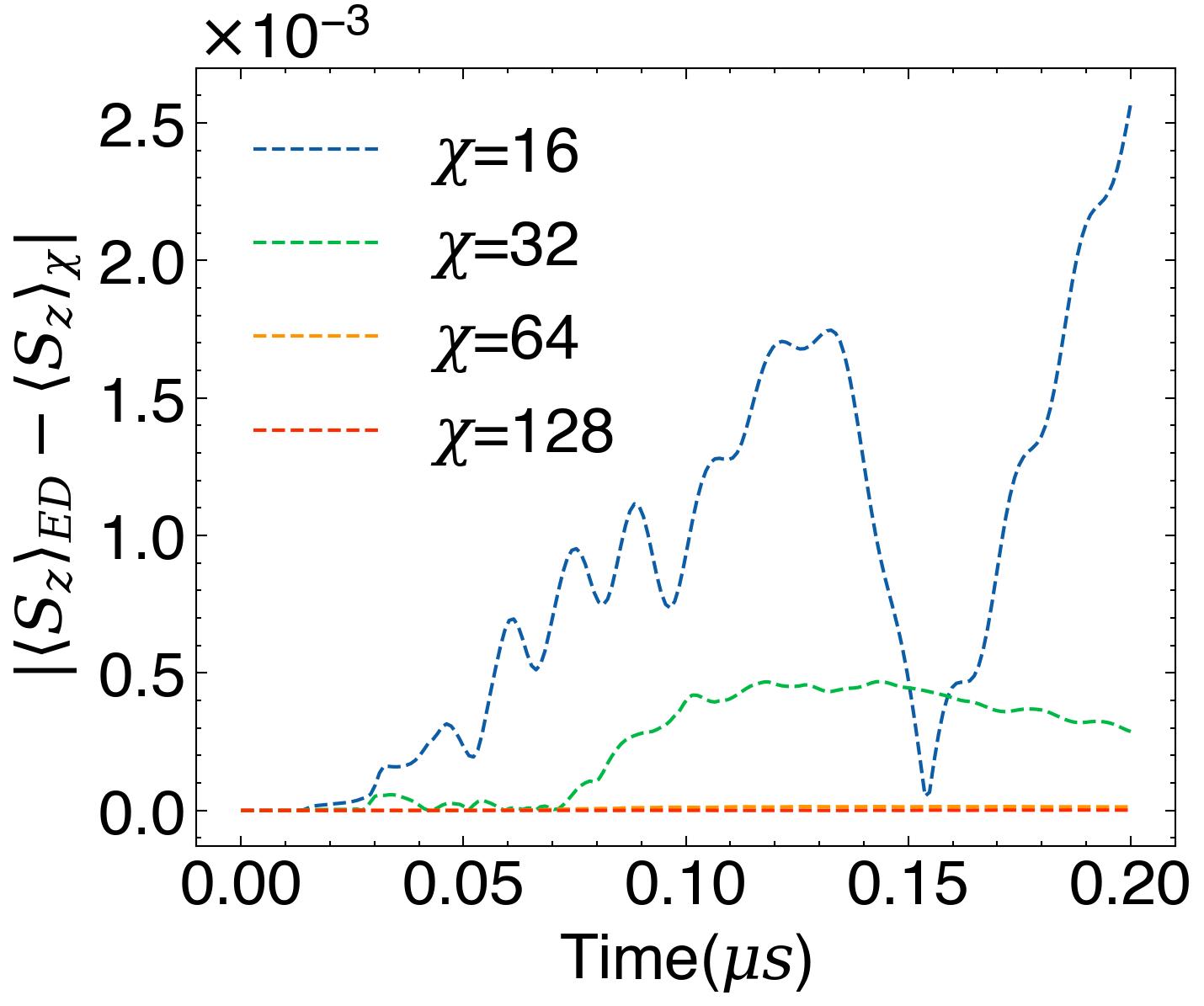}%
        \label{subfig:4NV_dis_a}%
    }\hfill
    \subfloat[]{%
        \includegraphics[width=.48\linewidth]{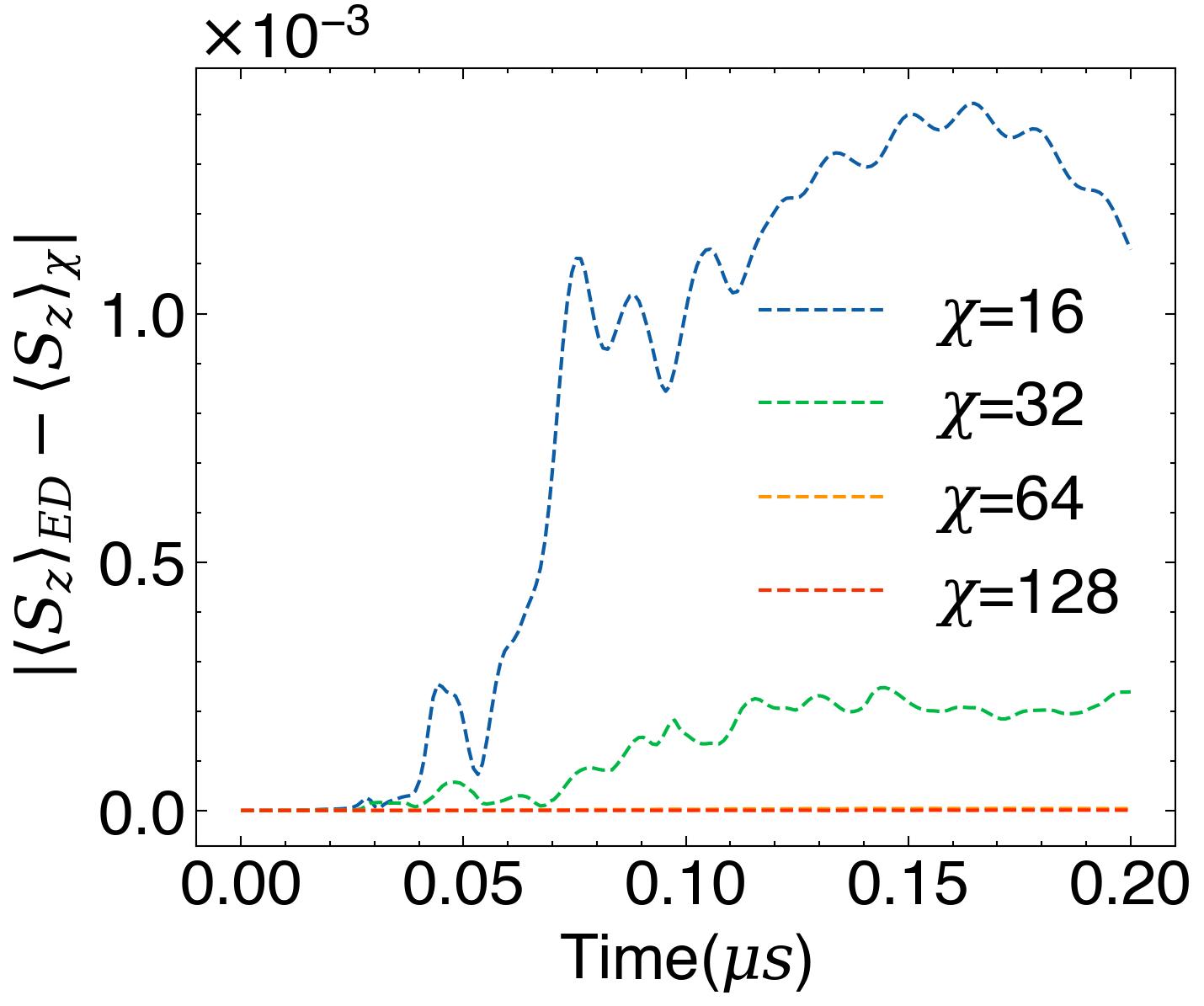}%
        \label{subfig:4NV_dis_b}%
    }\\
    \subfloat[]{%
        \includegraphics[width=.48\linewidth]{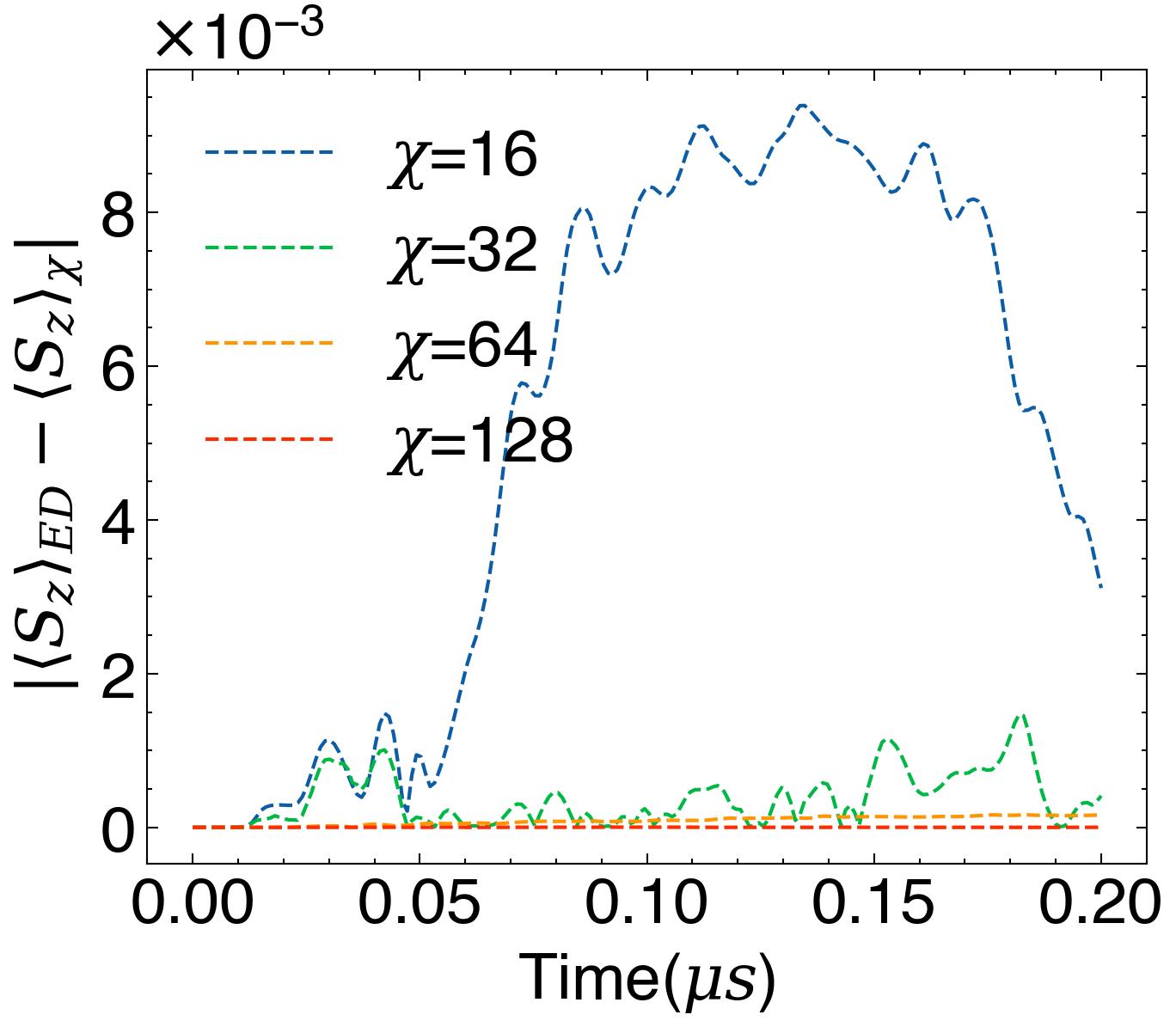}%
        \label{subfig:4NV_dis_c}%
    }\hfill
    \subfloat[]{%
        \includegraphics[width=.48\linewidth]{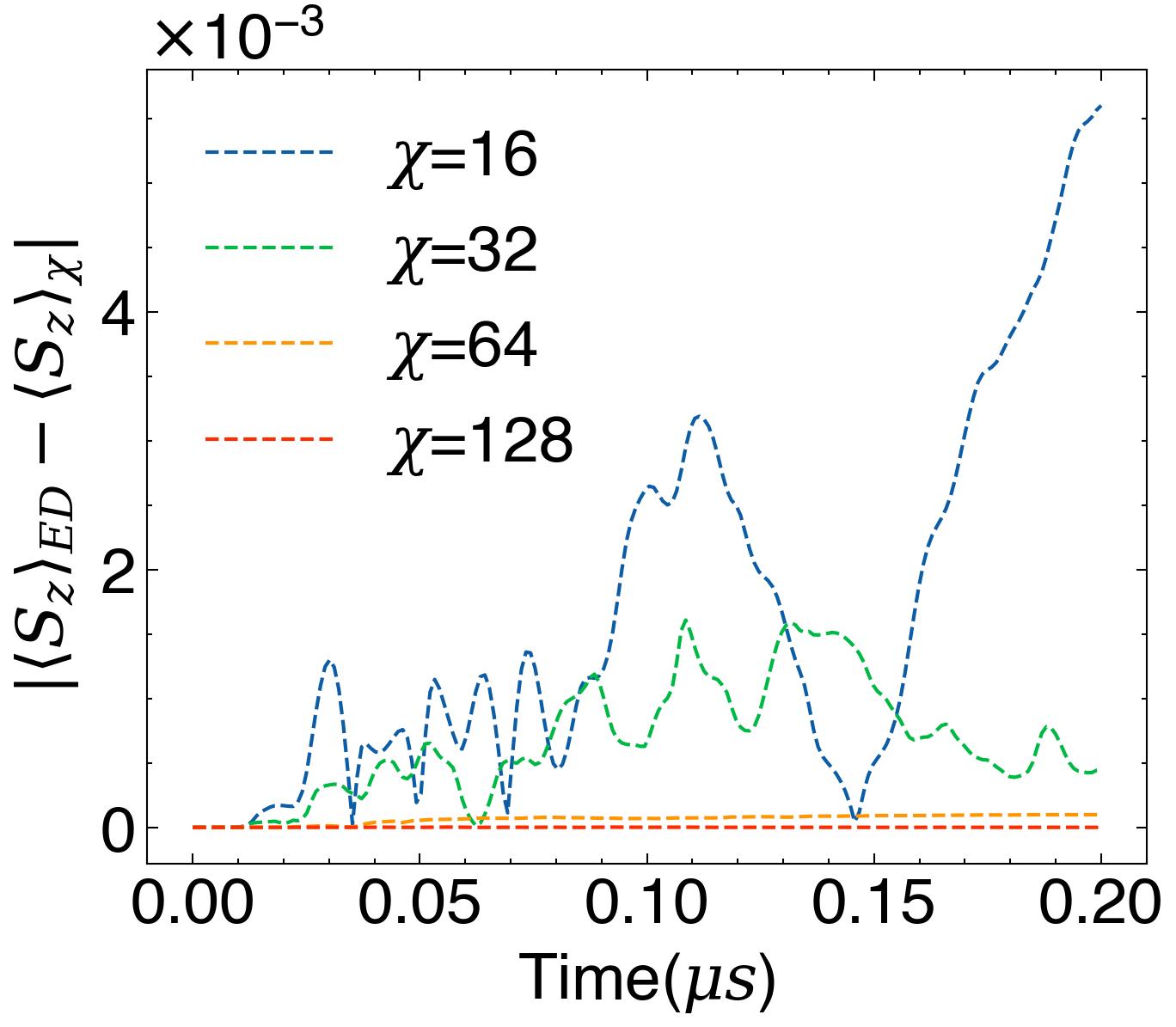}%
        \label{subfig:4NV_dis_d}%
    }
    \caption{Numerical errors compared to exact diagonalization as a function of maximum bond dimension for 4NVs. (\subref{subfig:4NV_dis_a}) $r=2.0$\,nm, $\gamma=1.0$, (\subref{subfig:4NV_dis_b}) $r=2.0$\,nm, $\gamma=5.0$, (\subref{subfig:4NV_dis_c}) $r=1.5$\,nm, $\gamma=1.0$, (\subref{subfig:4NV_dis_d}) $r=1.5$\,nm, $\gamma=5.0$.}
    \label{fig:4NV_dis}
\end{figure}

Next, we add dissipation by setting $\gamma_{i}=\gamma \neq
0$. So 
each spin has the same dissipation rate.  Numerical values of $\gamma$
are given in $(\mu \text{s})^{-1}$ units throughout. 
For the dissipation operator $\hat{L}_{i}$ we choose the dephasing
operator, $\hat{L}_{i}=\hat{S}_{z}$, relevant,
e.g., for
magnetic field noise. \cref{fig:4NV_dis} shows the absolute errors
compared to exact calculations of dissipative dynamics for different
$\chi_{\text{max}}$ and different values of $r$ and $\gamma$. As a result, the discrepancy of these dissipative
dynamics compared to the exact one is smaller than in the
non-dissipative case for the same $\chi_{\text{max}}$. When the growth of
singular values of bi-partition of the total state across the bond is
limited by
a larger $\gamma$, bond truncation becomes more effective
since it introduces less errors. This demonstrates the interplay
between the growth of entanglement entropy due to interactions and
dissipation that hinders it.  
To understand quantitatively how stronger interactions induce
entanglement entropy during time-evolution, we calculate the Operator
Entanglement Entropy (opEE) of the reduced density matrix determined
by cutting the system into two halves at the middle bond and tracing out
the other half. We use the von Neuman entropy defined as
\cite{murciano_more_2024}  
\begin{equation}
    S_{op} = -\sum_{i}^{\chi}(\lambda_{i})^{2}\log_{2}(\lambda_{i})^{2}, \label{eqn:opEE}
\end{equation}
where $\lambda_{i}$ are singular values of the cut bond with bond
dimension $\chi$ for the vectorized operator. Here the $\lambda_{i}$
are squared because they are singular values of a vectorized operator
$|\rho\rangle$. Note that in 
general opEE  does not give a direct measure of entanglement for mixed
states.
The opEE for a pure state is twice the value of its
standard 
entanglement entropy, $S_{op}(|\rho\rangle) =
2S(|\psi\rangle)$, when $\rho = |\psi\rangle \langle\psi|$ \cite{preisser_beltran_numerical_2023}.
Yet, the opEE provides insight into
how well the state can be approximated by an MPDO, indicating
simulation errors from a truncation \cite{preisser_comparing_2023}.

\begin{figure}[h]
    \subfloat[]{%
        \includegraphics[width=.48\linewidth]{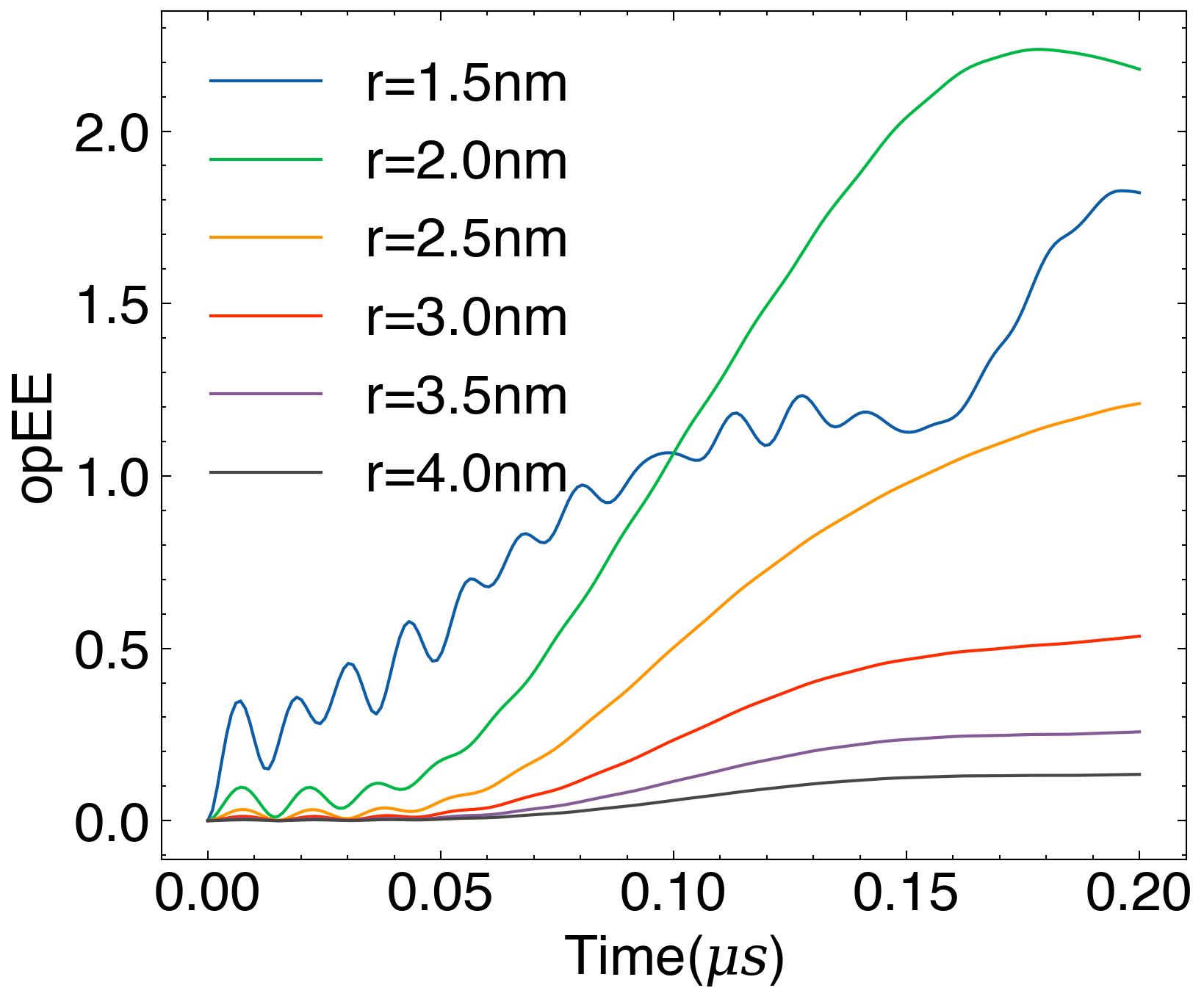}%
        \label{subfig:4NV_opEE}%
    }\hfill
    \subfloat[]{%
        \includegraphics[width=.48\linewidth]{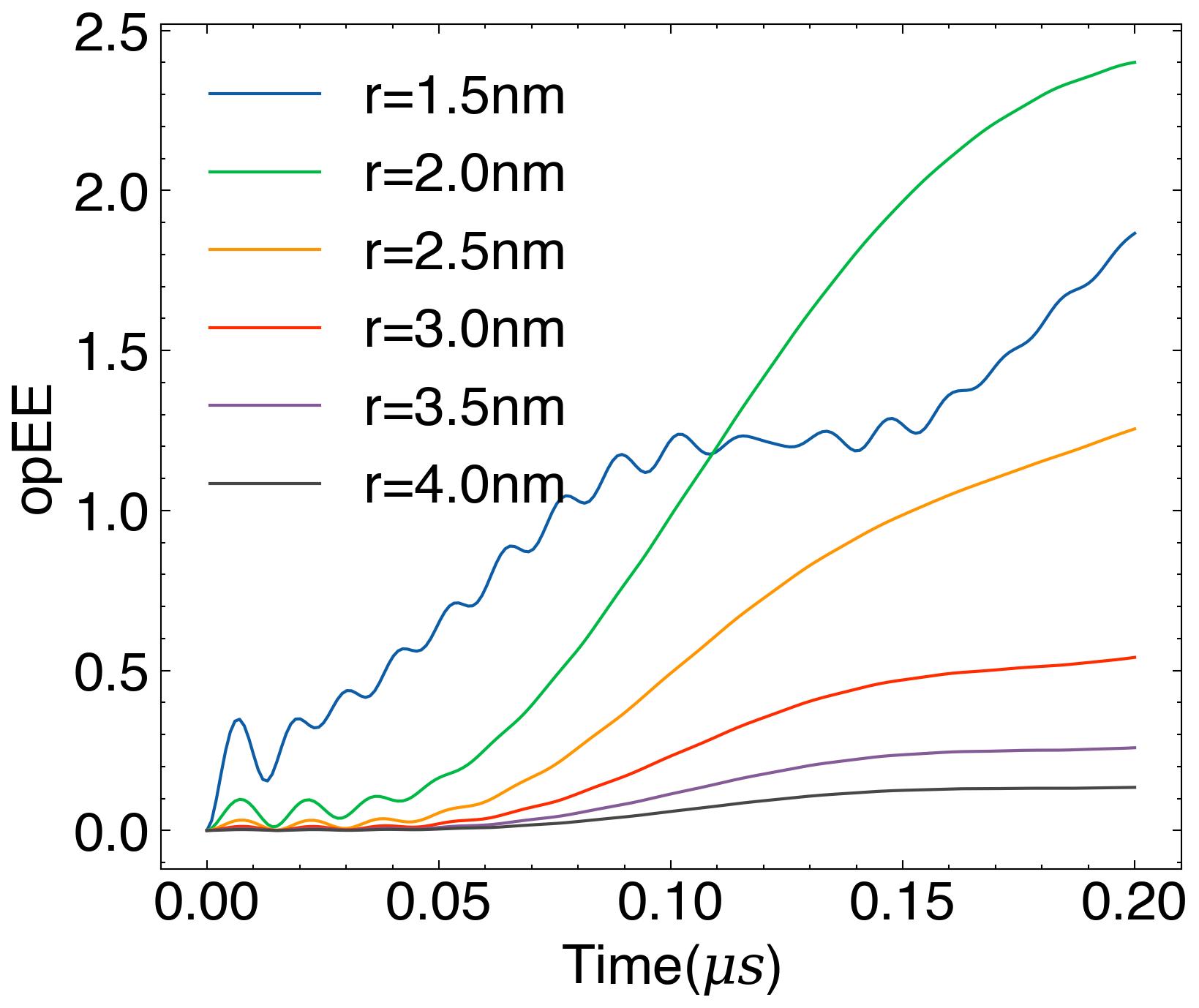}%
        \label{subfig:7NV_opEE}%
    } \\
    \subfloat[]{%
    \includegraphics[width=.48\linewidth]{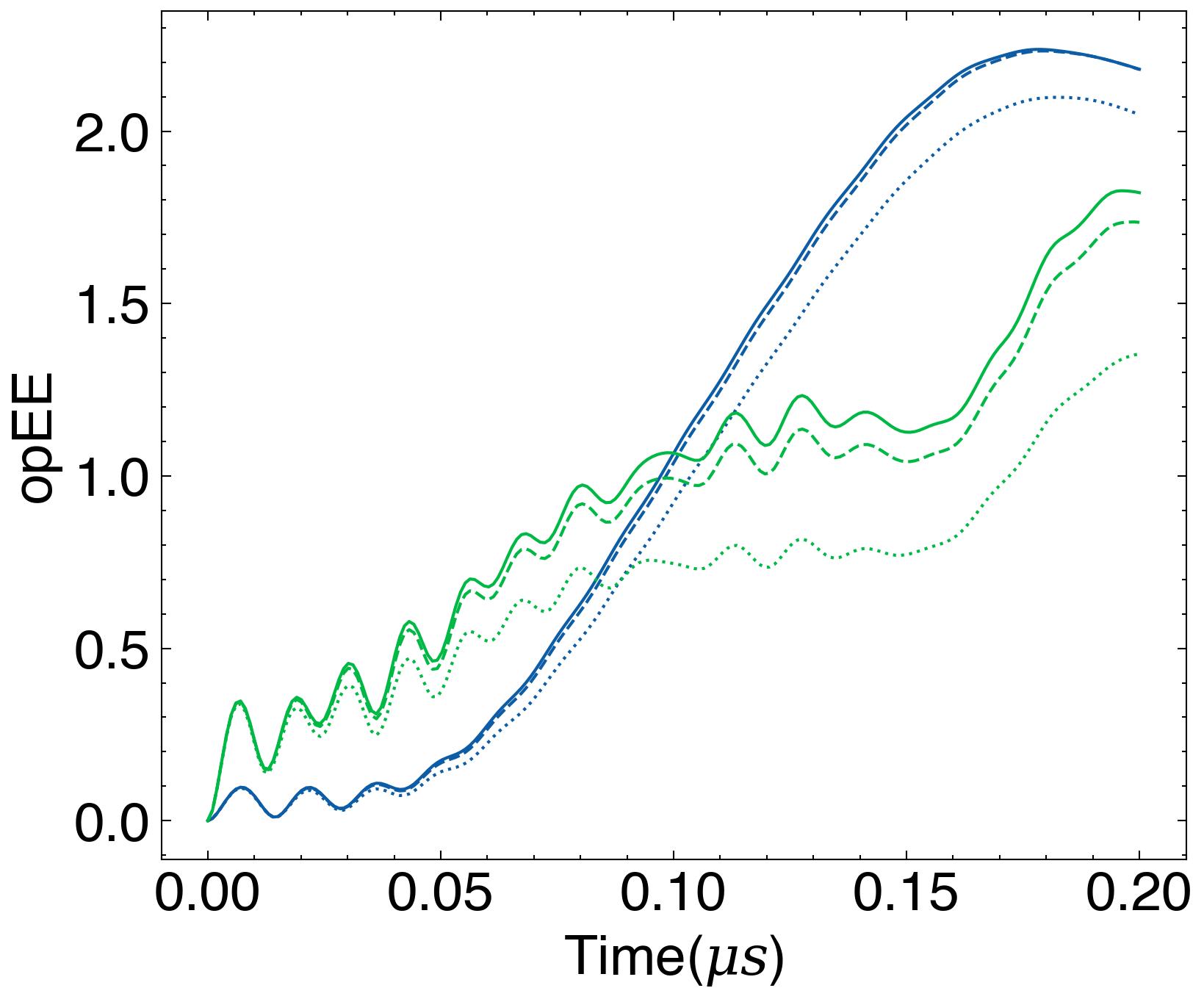}%
    \label{subfig:7NV_opEE_2nm}%
}\hfill
\subfloat[]{%
    \includegraphics[width=.48\linewidth]{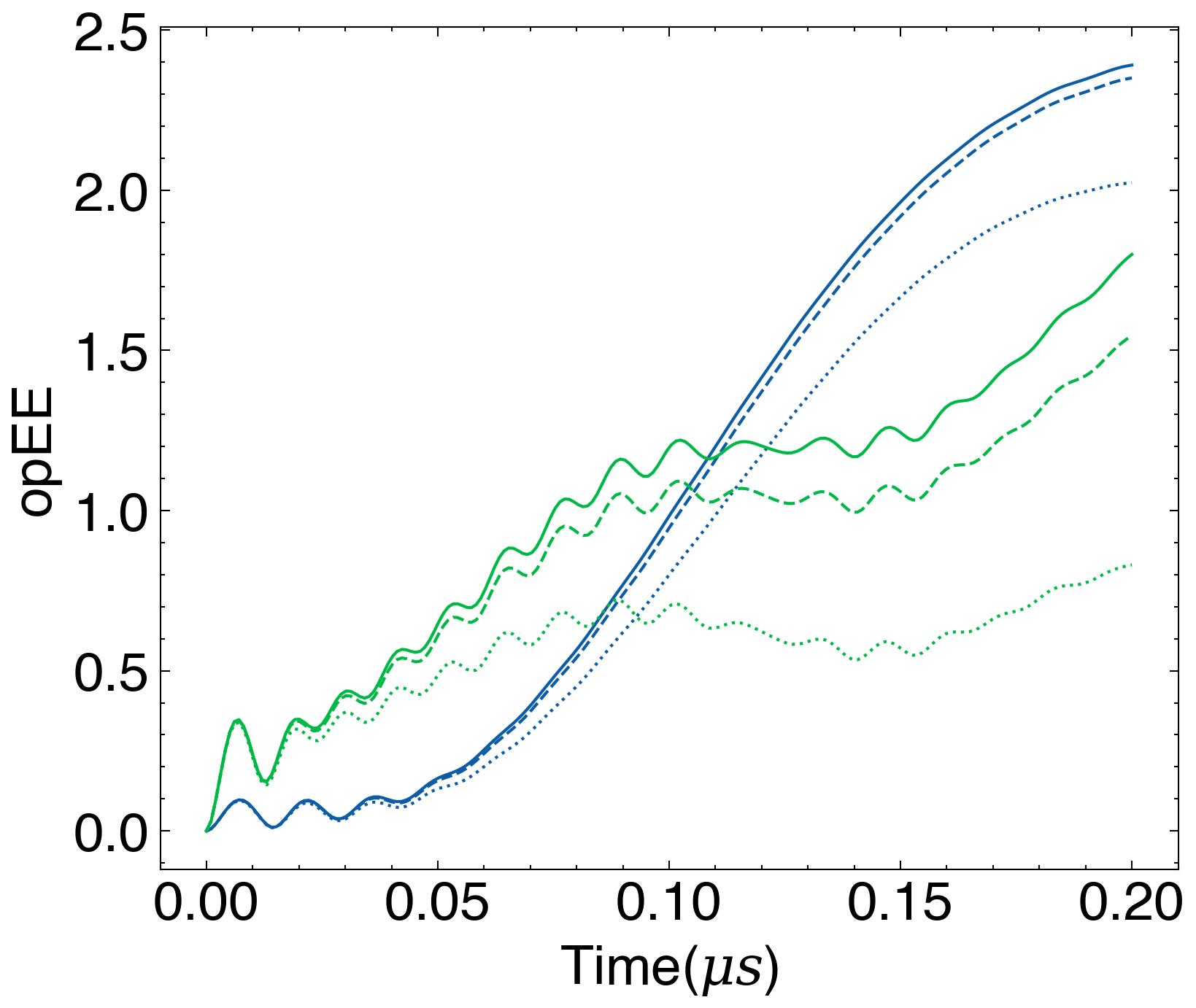}%
    \label{subfig:7NV_opEE_1.5nm}%
}
    \caption{Operator entanglement entropy (\cref{eqn:opEE}) of middle bond. (\subref{subfig:4NV_opEE}) and (\subref{subfig:7NV_opEE}), for $N=4$ and $7$ respectively, show trends of opEE to grow faster when interactions get stronger. Each lines has $\gamma=0$. (\subref{subfig:7NV_opEE_2nm}) $N=4$ and (\subref{subfig:7NV_opEE_1.5nm}) $N=7$, dissipation suppresses generation of opEE. Green: $r=1.5$\,nm, Blue: $r=2.0$\,nm. Line styles: solid ($\gamma = 0.0$), dashed ($\gamma = 1.0$), dotted ($\gamma = 5.0$).}
    \label{fig:opEE}
\end{figure}

After time-evolution, the middle bond index $\alpha_{(N/2 +1)}$ which
is the bond that equally separates the  MPDO into two halves has the largest dimension.
\cref{fig:opEE} shows the time evolution of opEE calculated from cutting
the middle bond for $N=4$ and $N=7$. \cref{subfig:4NV_opEE,subfig:7NV_opEE} are for $\gamma=0$, \cref{subfig:7NV_opEE_2nm,subfig:7NV_opEE_1.5nm} are for the different values of $\gamma$.
The opEE growth as function of time is accelerated as interactions
become stronger. This agrees with results of larger errors from
truncation shown earlier. Except at very strong interaction,
i.e.~for $r=1.5$\,nm, opEE grows at the beginning and saturates
rapidly.
For $r=1.5$\,nm, the initial growth is fastest, but continues to grow after short time interval of relatively stable value around 0.12$\mu s$. 
This behavior depends on the Rabi frequency $\Omega$. The doubling $\Omega$ from $(2\pi)2$MHz to $(2\pi)4$MHz, the initial rise of opEE becomes even faster but the plateau is reached around 0.12$\mu s$. See \cref{subfig:4NV_opEE_2Omega} in appendix. 

\subsection{QFI and sensitivity}
According to the Cram\'er-Rao bound, the sensitivity in the estimation of a
parameter $\theta$ using quantum probes is bounded by the inverse of
the Quantum Fisher Information (QFI), $(\delta \theta)^{2} \geq
\frac{1}{MF_{Q}}$ where $M$ is the number of independent measurements and $F_{Q}$ is QFI \cite{paris_quantum_2009}. The definition of QFI for a mixed state $\hat{\rho}$ is given by  
\begin{equation}
    F_{Q} = \Tr[\hat{\rho}\hat{L}^{2}], \label{eqn:qfi_definition}
\end{equation}
where $\hat{L}$ is the symmetric logarithmic derivative (SLD) that satisfies 
\begin{equation}
    \partial_{\theta}\hat{\rho} = \frac{1}{2}(\hat{L}\hat{\rho}+\hat{\rho}\hat{L}). \label{eqn:sld_definition}
\end{equation}

Hence, sensitivities can be improved by
repeating the measurements or using more probes. For $N$ independent
probes, we obtain the Standard Quantum Limit (SQL), $\delta \theta
\geq \mathcal{O}(1/\sqrt{N})$. However, the QFI can be increased when
the probes are highly entangled. For example, under evolution with a
pure Zeeman term, and in the absence of decoherence and dissipation,
probes prepared in a GHZ state achieve optimal sensitivity for
magnetic field measurement that follows the "Heisenberg limit" (HL),
$\delta \theta \geq \mathcal{O}(1/N)$
\cite{giovannetti_quantum-enhanced_2004}. 

Thus, since interactions are necessary for the creation of
entanglement, dense NV ensembles harbor the potential for higher sensitivity
compared to non-interacting NVs. However, the lack of permutational
symmetry leads to the population of other irreducible representations
of SU(2) starting from the one with maximum spin, i.e.~on average the
total spin decays and sensitivity is reduced. Optimal control is
therefore necessary to harvest the entanglement from strong
interactions for achieving higher sensitivity. 

\subsection{Dynamics of QFI}

To quantify the sensitivity of the ensemble of interacting NVs to a uniform magnetic field $B_{z}$, we compute the Quantum Fisher Information of the mixed state during time evolution. Instead of direct calculation using \cref{eqn:qfi_definition}, we use the tensor network approach for calculating QFI as in \cite{chabuda_tensor-network_2020}. Note that, unlike in the original work \cite{chabuda_tensor-network_2020}, we only optimize the SLD but not the input state $\rho$. In short, we iteratively search for the SLD that satisfies a definition of QFI:
\begin{equation}
    F(\rho_{\phi},L)=\underset{L}{\text{sup}}[2\text{Tr}(\rho'_{\phi}L) - \text{Tr}(\rho_{\phi}L^{2})] \label{qfi}.
\end{equation}
At the beginning, we randomly create the MPO approximation of an SLD, $L$, given by 
\begin{equation}
    L = \sum_{jk} \text{Tr}(S[1]_{k_{1}}^{j_{1}}...S[n]_{k_{n}}^{j_{n}})|\textbf{j}\rangle\langle\textbf{k}|.
\end{equation}
Here $S[l]_{k_{l}}^{j_{l}}$ is a Hermitian matrix. In searching for an optimal $L$ we locally update $(S[1]_{k_{1}}^{j_{1}},...,S[n]_{k_{n}}^{j_{n}})$ by sweeping from $S[1]$ to $S[n]$ and back to $S[1]$ again until $F$ has converged. For example, when updating $S[l]_{k_{l}}^{j_{l}}$, all other tensors $S_{k}^{j}$ are fixed and combined. Then after contractions, \cref{qfi} becomes 
\begin{equation}
    F(\rho,L) = 2\sum_{\alpha}b_{\alpha}S[l]_{\alpha} - 2\sum_{\alpha\beta}S[l]_{\alpha}A_{\alpha\beta}S[l]_{\beta}. \label{qfi_2}
\end{equation}
Here, $b$ and $A$ are a vector and a matrix resulting from contracting
the fixed tensors and combining the un-contracted indices. The
diagrammatic explanation can be found on page 8 of
\cite{chabuda_tensor-network_2020}. Taking the derivative with respect
to $S[l]$, $\partial F(\rho,L)/\partial S[L] = 0$, \cref{qfi_2} yields
$\frac{1}{2}(A+A^{T})|S[l]\rangle = |b\rangle$.  A solution to this
equation provides a local extremum for the QFI.
Since this local update approach tends to get stuck at local extrema, several repetitions with different initial $L$ are needed and can be computational expensive for large $N$. 

\begin{figure}[h]
    \subfloat[]{%
        \includegraphics[width=.48\linewidth]{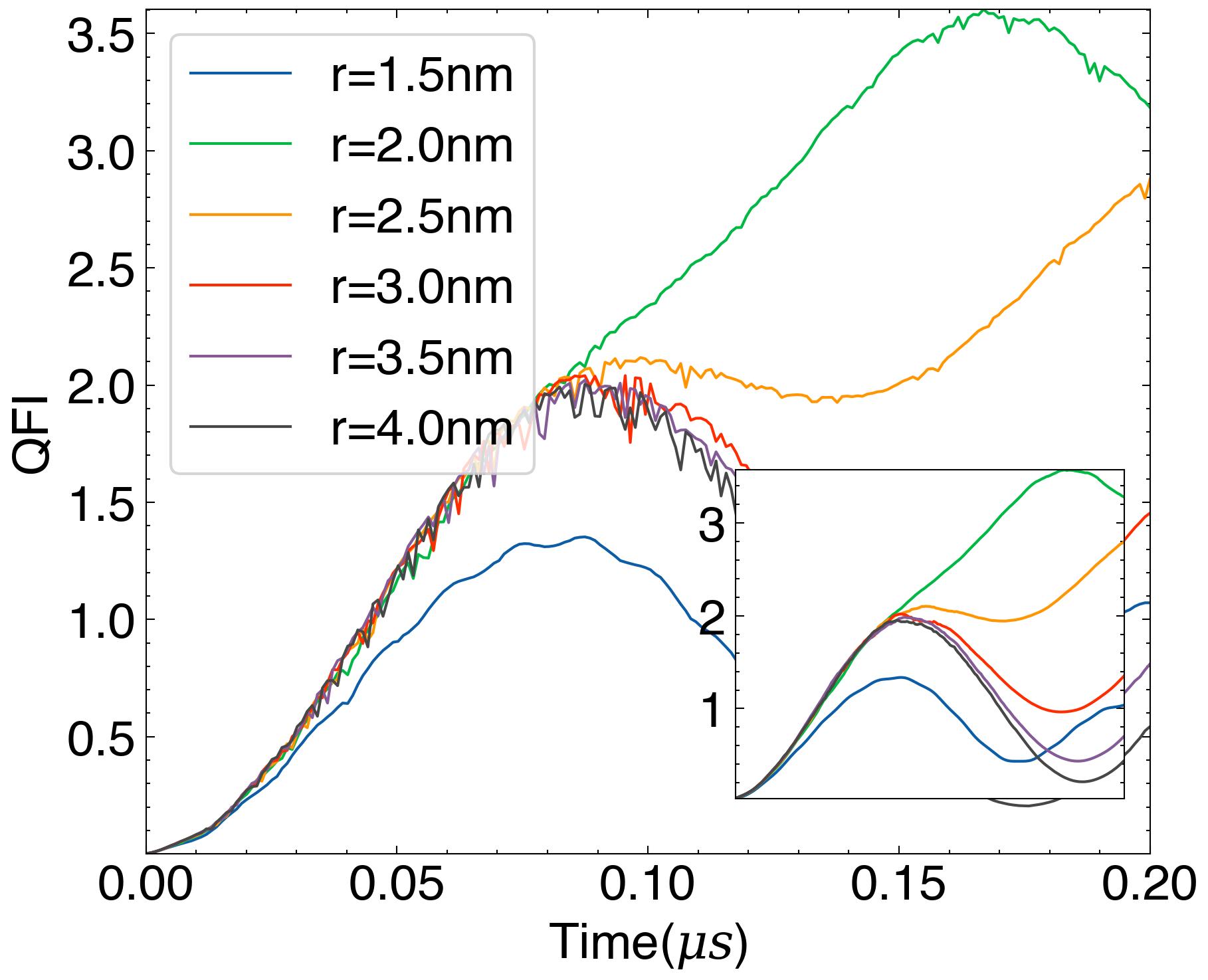}%
        \label{subfig:2NV_qfi}%
    }\hfill
    \subfloat[]{%
        \includegraphics[width=.48\linewidth]{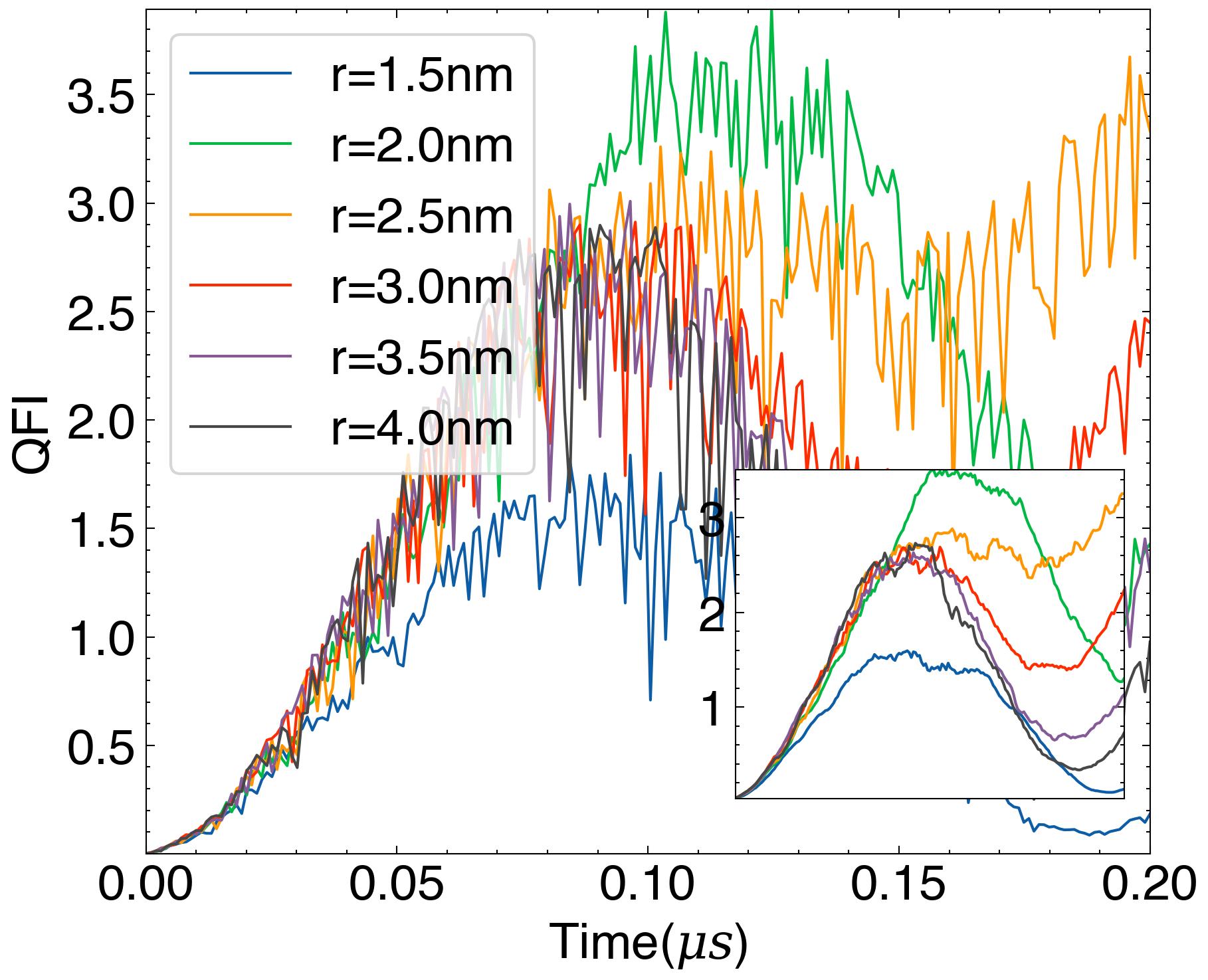}%
        \label{subfig:3NV_qfi}%
    }
    \subfloat[]{%
        \includegraphics[width=.48\linewidth]{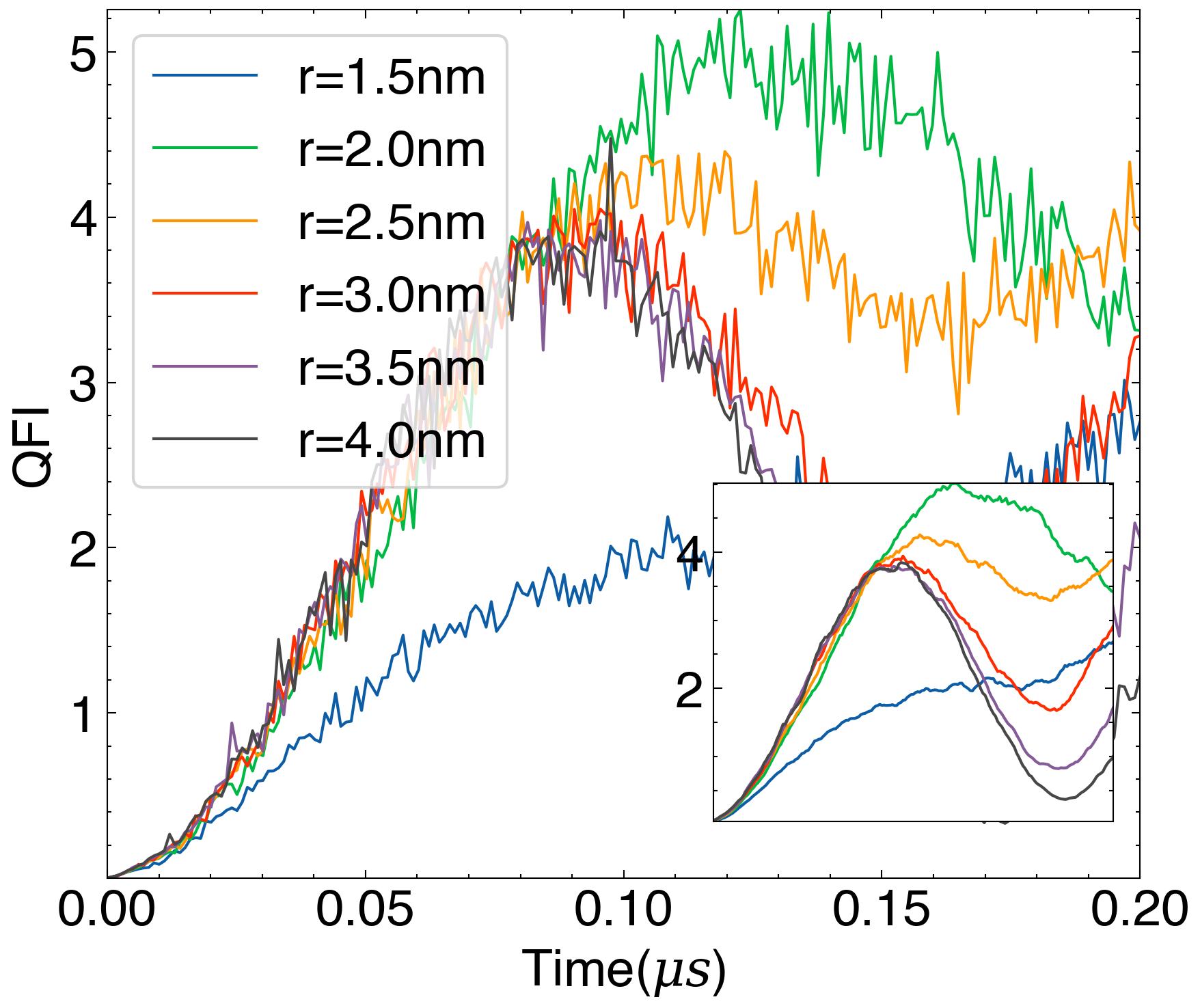}%
        \label{subfig:4NV_qfi}%
    }
    \caption{Dynamics of QFI with Rabi frequency = $\Omega= 2\pi \cross 2.00$ MHz. All plots are the best value computed from 10 different SLD initializations. The insets contain moving averages over 10 time steps. (\subref{subfig:2NV_qfi}) $N=2$, (\subref{subfig:3NV_qfi}) $N=3$, (\subref{subfig:3NV_qfi}) $N=3$.}
    \label{fig:qfi}
\end{figure}

\cref{fig:qfi} shows the dynamics of the QFI for a non-dissipative
system with small size and different $r$. At each time step, we find
the optimal SLD from 10 independent realizations and select the one
with the largest QFI. Furthermore, we utilize the optimal SLD obtained
from the previous time step as the initialization for the current time
step. Based on our experience, although not guaranteed, implementing
such a strategy 
can
lead to a smoother result
for the QFI.

\begin{figure}[h]
    \subfloat[]{%
        \includegraphics[width=.48\linewidth]{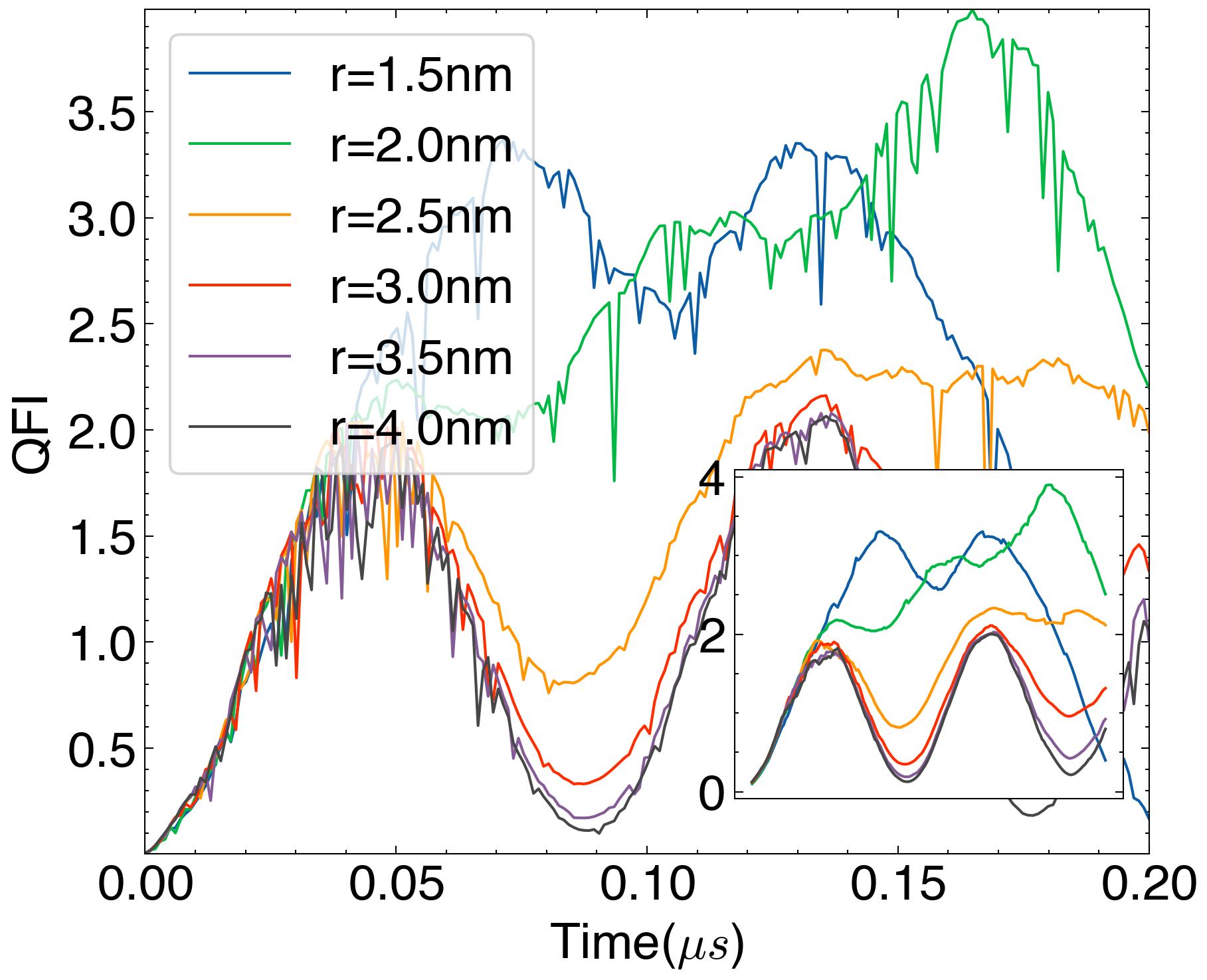}%
        \label{subfig:2NV_qfi_2omega}%
    }\hfill
    \subfloat[]{%
        \includegraphics[width=.48\linewidth]{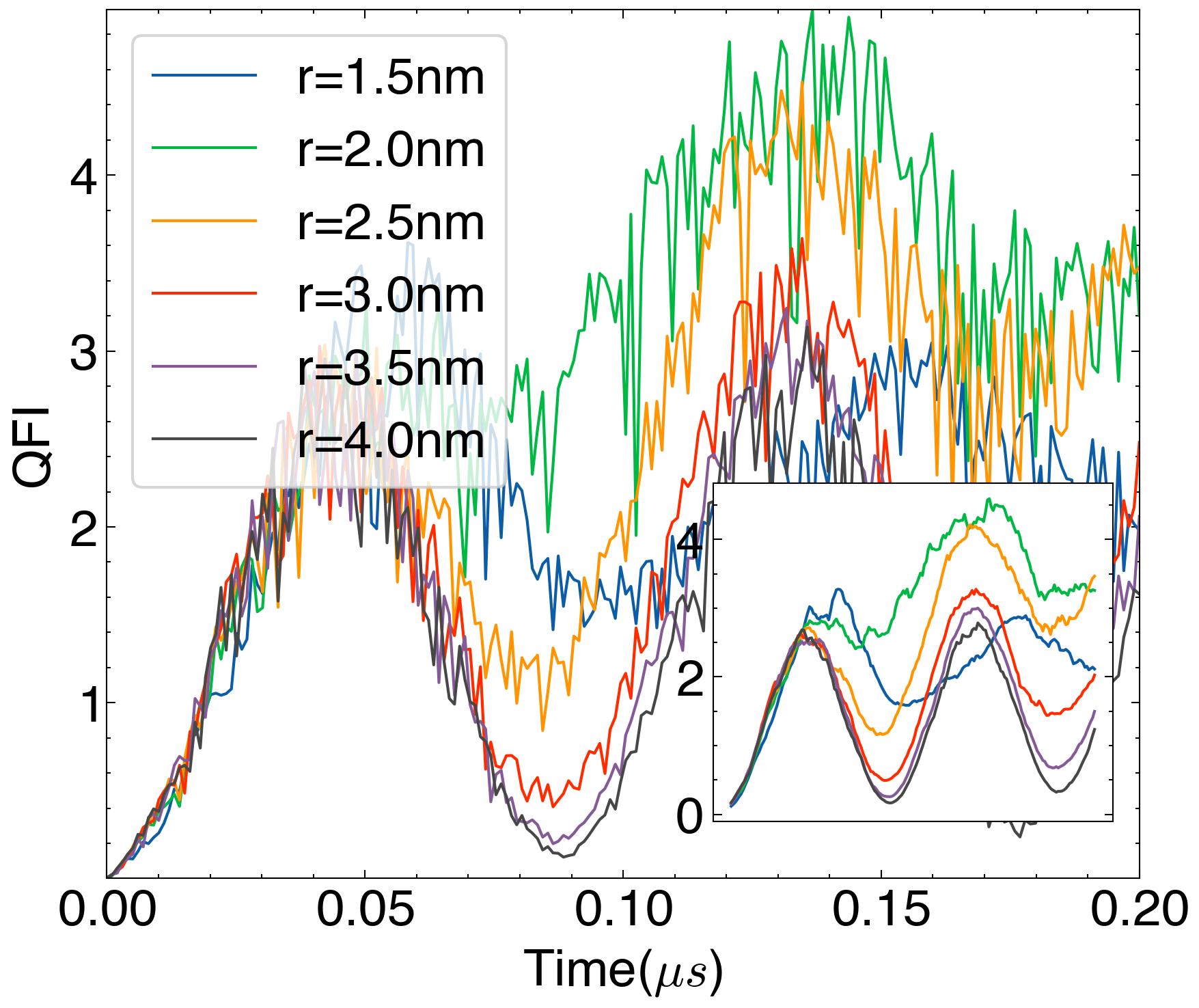}%
        \label{subfig:3NV_qfi_2omega}%
    }
    \subfloat[]{%
        \includegraphics[width=.48\linewidth]{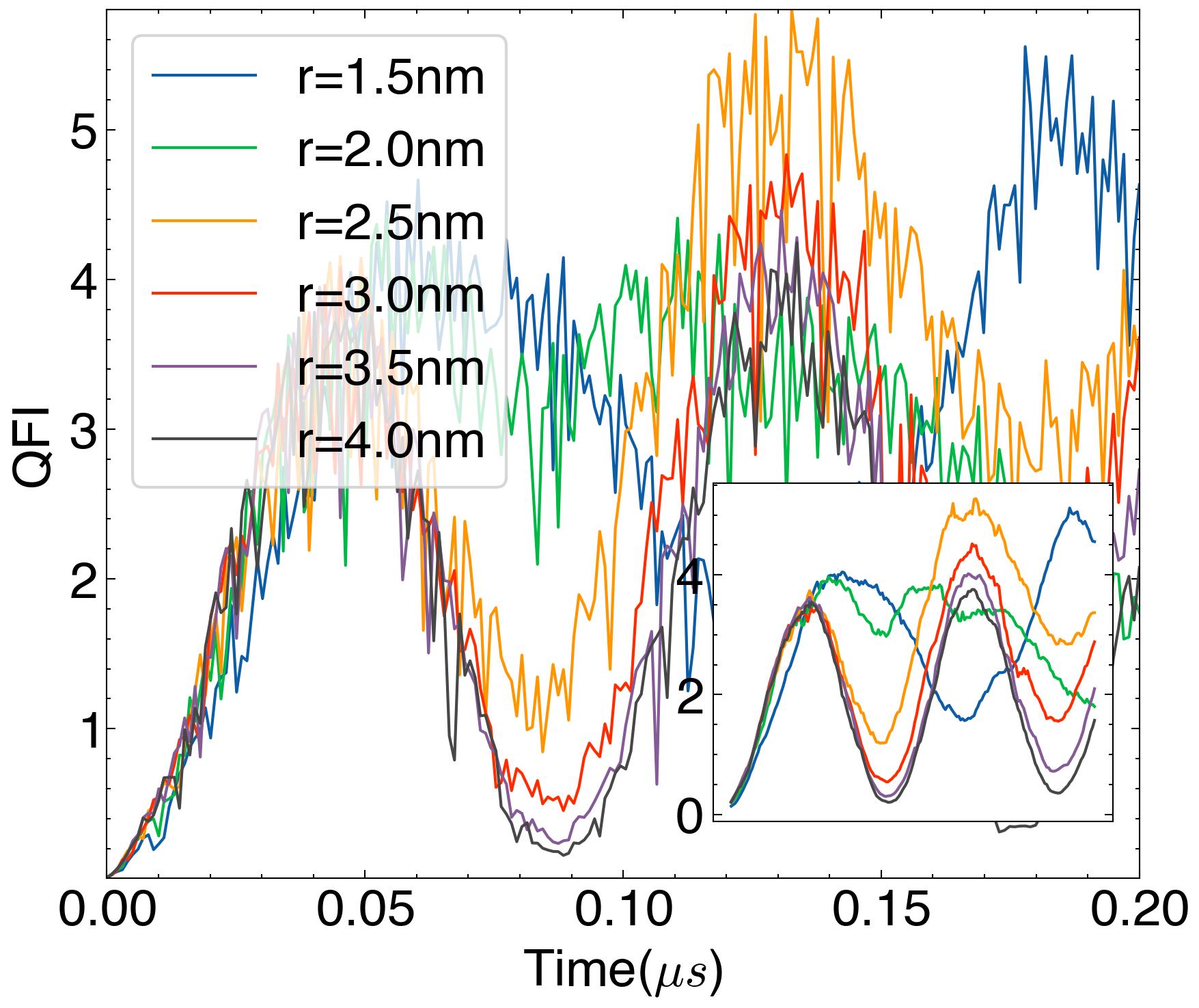}%
        \label{subfig:4NV_qfi_2omega}%
    }
    \caption{Dynamics of QFI when the Rabi frequency is doubled from \cref{fig:qfi} to $2\Omega=(2\pi)4.00$\,MHz. An increase of the maximal value of the QFI compared to the case with Rabi frequency $\Omega =2\pi \times 2.00$ MHz can be observed for $r=1.5$nm. (\subref{subfig:2NV_qfi_2omega}) $N=2$, (\subref{subfig:3NV_qfi_2omega}) $N=3$, (\subref{subfig:4NV_qfi_2omega}) $N=4$.
    }
    \label{fig:qfi_2Omega}
\end{figure}

According to the plots,  
the QFI of $r=4.0$\,nm nearly reaches values of $N$ non-interacting probes, denoted
as $F_{|+1\rangle^{\otimes N}}=N$. The QFI increases if the interactions are stronger, especially when $r\leq2.5$\,nm, leading to a QFI that can be greater than $N$.
This suggests that the sensitivity for the probes is enhanced by the
entanglement that is created by interaction. Still, as
for
opEE, the QFI remains relatively small when $r=1.5$\,nm, i.e.~for very strong interaction. 
We find that this is due to an interaction strength of the nearest NV pairs ($C_{dip,(i,i+1)}=2\pi\times 15.41$\,MHz)
that becomes substantially larger than the Rabi frequency ($\Omega=2\pi\times 2.00$\,MHz). 
As shown \cref{fig:qfi_2Omega}, the QFI can be further increased also for $r=1.5$\,nm by increasing the Rabi frequency.

\section{Conclusion}

An ensemble of NV centers is a promising quantum metrology
platform. While adding more independent spins
improves the sensitivity, the probe can
in principle also reach greater sensitivity due to entanglement
created by interactions.
Also, having a dense ensemble means that the total size of the probe can be
smaller, resulting in better spatial resolution. However, without external
control of the entanglement and the states generated, strong
dipole-dipole interactions due to small separations between NV centers
within a dense ensemble are usually detrimental for the sensitivity as
they lead to a rapid decay of the total spin. 

To investigate a system of strongly interacting NVs, we simulated the master equation with dissipation using MPDO. We benchmarked the $W^{II}$ and the time-dependent variational principle (TDVP) algorithms against exact numerical diagonalization method for time evolution with a long-range interaction model.
We found the $W^{II}$ algorithm to have bigger errors in
the imaginary part and to be less stable in a very small separation
regime than TDVP. Then, with TDVP for small NV separations, 
we investigated the effect
of finite maximum bond dimensions on the simulation accuracy. We again determined
the accuracy as a function of bond dimensions by comparing
the obtained states to those from
exact diagonalization. Stronger
interactions result in larger errors, suggesting the need for larger
bonds. For dissipative dynamics,
we found
higher accuracy,
which can be attributed to slower growth
of operator entanglement entropy.
These results indicate that the
tensor network method can be suitable
for simulating
an open system.
However, we find an exception for $r=1.5$\,nm, where the errors can be bigger even though the operator entanglement entropy is smaller than for $r=2.0$\,nm.

For applications in quantum sensing,
we used the approach of
locally optimizing approximations of the SLD
to obtain the quantum Fisher information. The
QFI of driven ensembles shows that strong interaction can create
entanglement-enhanced sensitivity 
compared to
independent spins.
Our results are based on quantum Fisher information, which gives the ultimate achievable sensitivity optimized over all possible POVM measurements and unbiased estimator functions of the parameter.
Additional investigation will have to show to what extent this optimal sensitivity can be reached in specific sensing protocols.

Finally, we observed that the boost in sensitivity can diminish when
the interaction becomes significantly
larger than the Rabi frequency.
In this situation, it is necessary to utilize microwaves with higher
intensities
for driving in order to increase the sensitivity.

\appendix
\section{Pure state entanglement entropy for a vector and an operator.}
Any pure state has a Schmidt decomposition
$
    |\psi_{AB}\rangle = \sum_{i}^{r}\sqrt{\lambda}_{i} |i\rangle_{A}\otimes |i\rangle_{B} $
    with Schmidt coefficients $\sqrt{\lambda}_{i} > 0$ and $\sum_{i}^{r}\lambda_{i} = 1$.
Then we can define the reduced density matrix for a bipartite system by partial trace over the other parts:
$\rho_{A} = \Trace_{B}[\rho_{AB}] = \sum_{i}^{r} \lambda_{i} |i\rangle_{A}\langle i|_{A}$ 
and
$\rho_{B} = \Trace_{A}[\rho_{AB}] = \sum_{j}^{r} \lambda_{j} |j\rangle_{B}\langle j|_{B}.$
The density operator of the total system reads $\rho_{AB} = |\psi_{AB}\rangle\langle\psi_{AB}| = \sum_{ij}^{r,r} \sqrt{\lambda_{i}}\sqrt{\lambda_{j}}  (|i\rangle_{A} \otimes |i\rangle_{B}) (\langle j|_{A} \otimes \langle j|_{B})$.
The entanglement entropy for reduced density operators is defined as
\begin{align}
    S(\rho_{A}) = -\Trace[\rho_{A}\log(\rho_{A})] 
        = -\sum_{i=1}^{r}\lambda_{i}\log{\lambda_{i}},\label{standard_S}
\end{align}
where $S(\rho_{A}) = S(\rho_{B})$. To find the operator entanglement entropy (opEE), we vectorize $\rho_{AB}\rightarrow|\rho_{AB}\rangle\rangle$,
\begin{align}
    |\rho_{AB}\rangle\rangle &= |\psi_{AB}\rangle|\psi_{AB}\rangle \\
    &= \sum_{\mu}^{r^{2}}\Lambda_{\mu} |i_{A}i_{B};j_{A}j_{B}\rangle.
\end{align}
Here $\Lambda_{\mu} = \sqrt{\lambda_{i}}\sqrt{\lambda_{j}}$ and 
$\mu=(i,j)$. The super density operator, denoted by $\rho^{\sharp}$, can be created using an outer product:
\begin{align}
    \rho^{\sharp}_{AB} &=  |\rho_{AB}\rangle \langle\rho_{AB}|\\
    &= \sum_{\substack{\mu=(i,j),\\ \nu=(k,l)} }^{r^{2},r^{2}} \Lambda_{\mu}\Lambda_{\nu} |i_{A}i_{B};j_{A}j_{B}\rangle \langle k_{A}k_{B};l_{A}l_{B}|.
\end{align}
Similar to $\rho_{A}(\rho_{B})$, we can also have $\rho_{A}^{\sharp}(\rho_{B}^{\sharp})$; e.g.~
$
    \rho_{A}^{\sharp} = \sum_{\mu}^{r^{2}} \Lambda_{\mu}^{2} |i_{A};j_{A}\rangle \langle i_{A};j_{A}|, \label{sup_rho_A}
$
Then we can find the opEE,
\begin{align}
    S_{OP}(\rho_{A}^{\sharp}) &= -\left[\sum_{\mu} \Lambda_{\mu}^{2} \log(\Lambda_{\mu}^{2})\right] \\
                            &= -\left[\sum_{i,j} (\lambda_{i}\lambda_{j}) \log(\lambda_{i}\lambda_{j})\right] \\
                            &= -\left[\sum_{i,j} (\lambda_{i}\lambda_{j}) \left( \log(\lambda_{i}) + \log(\lambda_{j})\right)\right] \\
                            &= -\left[\sum_{i} (\lambda_{i})\log(\lambda_{i})\sum_{j}(\lambda_{j}) + \sum_{i}(\lambda_{i})\sum_{j}(\lambda_{j})\log(\lambda_{j})\right] \\
                            &= -\left[2\sum_{i}\lambda_{i}\log(\lambda_{i})\right]  \\
                            &= 2S(\rho_{A}).
\end{align}

\section{opEE for Rabi frequency $2\Omega$}
The operator entanglement entropy when the Rabi frequency is doubled from $\Omega=(2\pi)2.00$\,MHz, hence $2\Omega = 2\pi\times 4.00$\,MHz.
\begin{figure}[h]
    \subfloat[]{%
        \includegraphics[width=.48\linewidth]{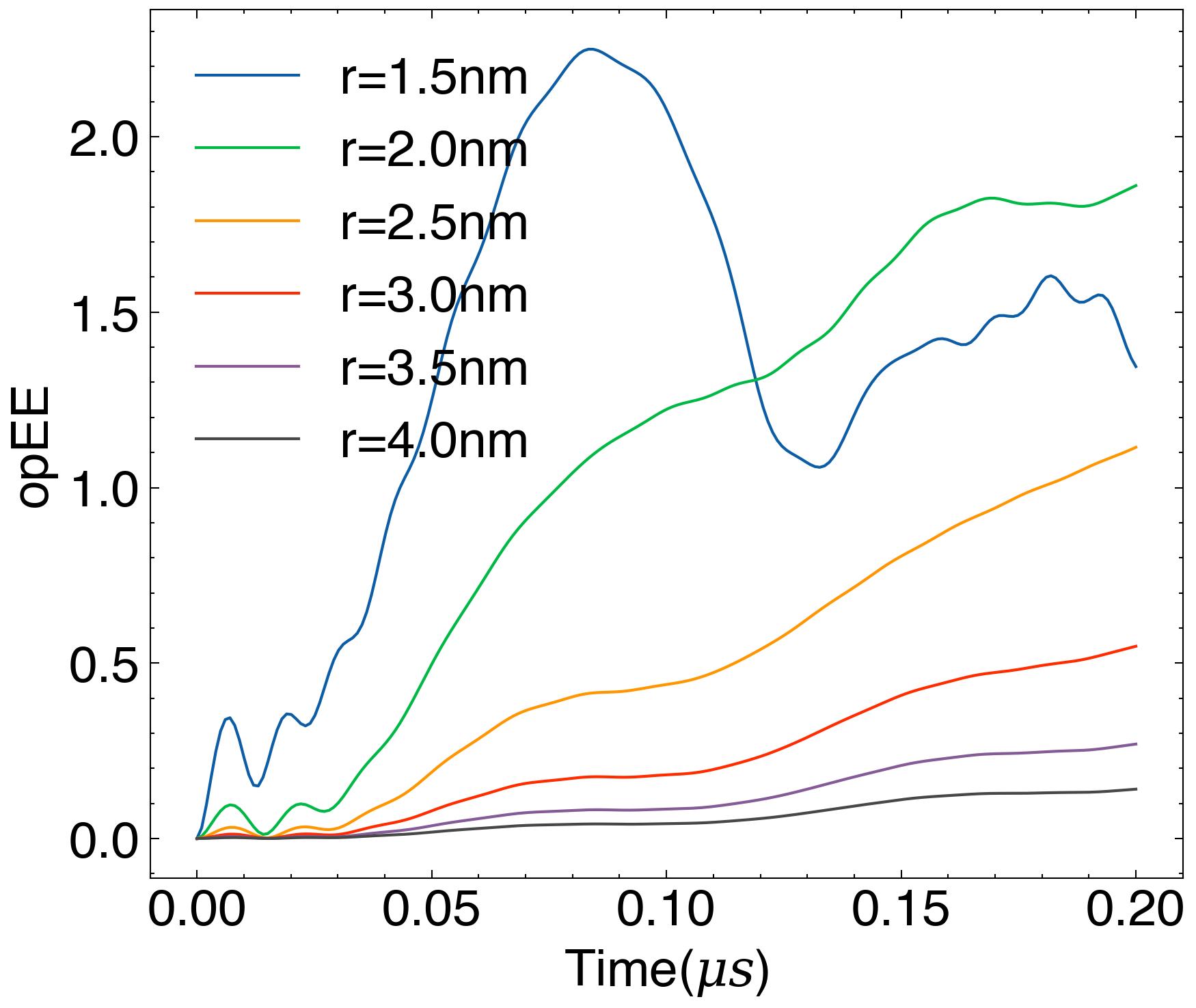}%
        \label{subfig:4NV_opEE_2Omega}%
    }\hfill
    \subfloat[]{%
        \includegraphics[width=.48\linewidth]{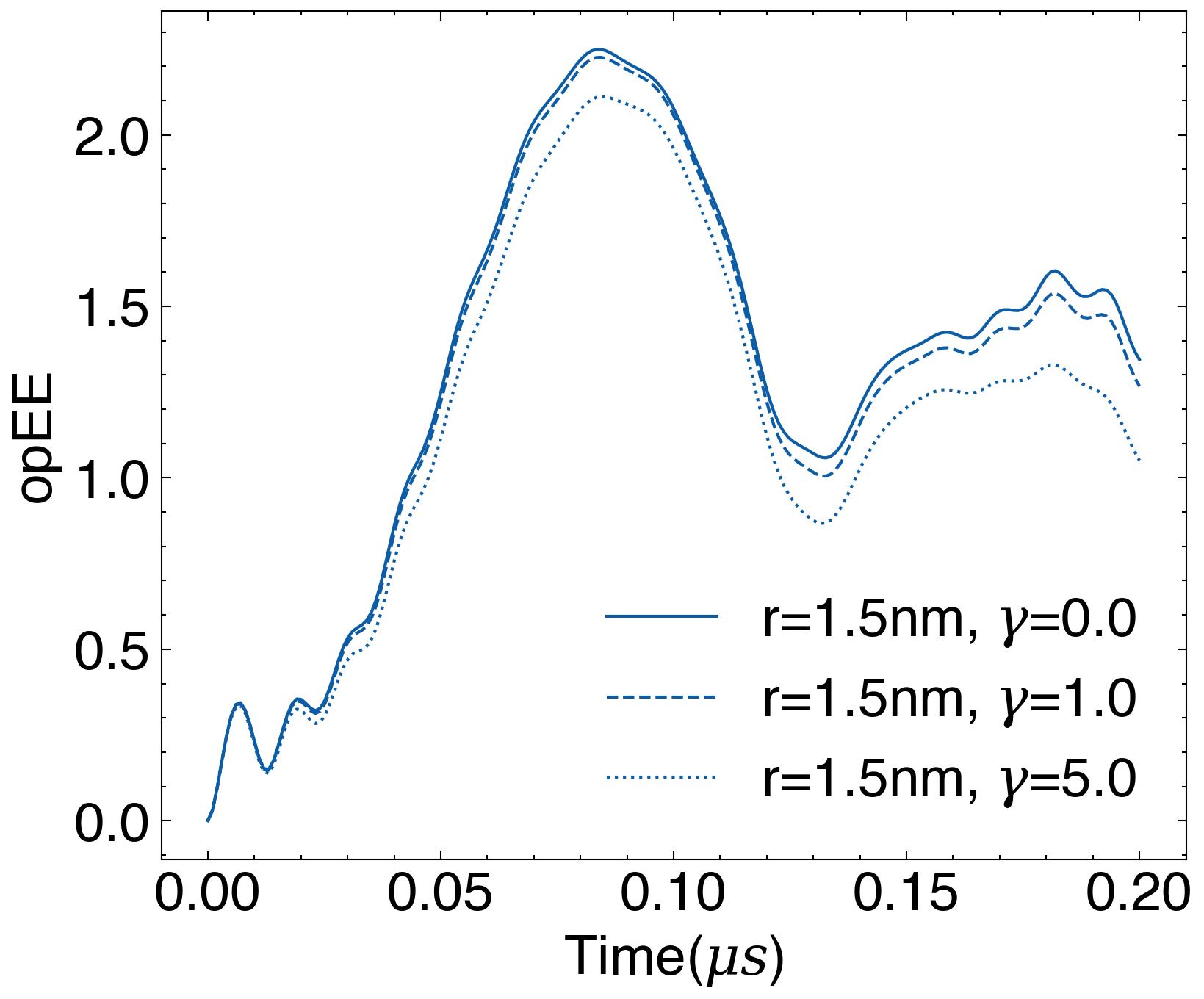}%
        \label{subfig:4NV_opEE_2Omega_Lz1}%
    }
    \caption{Operator entanglement entropy of middle bond. Rabi frequency = $2\Omega$.}
    \label{fig:opEE_2Omega}
\end{figure}
\begin{acknowledgments}
  This work was supported by the Baden W\"urttemberg Stiftung, project CDINQUA.  Tensor network codes are modified and 
  implemented
  algorithms from TenPy \cite{tenpy}.
\end{acknowledgments}

\bibliography{bib}

\end{document}